\documentclass[a4paper,11pt]{article}

\usepackage{amsmath}
\usepackage{amssymb}
\usepackage{amsthm}
\usepackage{indentfirst}
\usepackage{float}
\usepackage{graphicx}
\usepackage[utf8]{inputenc}
\usepackage[dvipsnames]{xcolor}
\usepackage{tabularx}
\usepackage{caption}
\usepackage{subcaption}
\usepackage{gensymb}
\usepackage[normalem]{ulem}
\usepackage[title]{appendix}
\usepackage{diagbox}

\usepackage{jheppub}
\hypersetup{colorlinks=true, linkcolor=blue, urlcolor=blue, citecolor=blue, linktocpage=true}

\begin{document}

\title{Methods of machine learning for the analysis of cosmic rays mass composition with the KASCADE experiment data}

\author[a]{M. Yu. Kuznetsov}
\author[a, b, c]{N. A. Petrov}
\author[a, b, d]{I. A. Plokhikh}
\author[a]{V. V. Sotnikov}

\affiliation[a]{Institute for Nuclear Research of the Russian Academy of Sciences,
    117312, Moscow, Russia}
    
\affiliation[b]{Novosibirsk State University,
    630090, Novosibirsk, Russia}
  
\affiliation[c]{Budker Institute of Nuclear Physics, SB RAS,
    630090, Novosibirsk, Russia}

\affiliation[d]{Institute of Thermophysics, SB RAS,
    630090, Novosibirsk, Russia}

\emailAdd{mkuzn@inr.ac.ru}

\abstract{We study the problem of reconstruction of high-energy cosmic rays mass composition from the experimental data of extensive air showers. We develop several machine learning methods for the reconstruction of energy spectra of separate primary nuclei at energies 1-100~PeV, using the public data and Monte-Carlo simulations of the KASCADE experiment from the KCDC platform. We estimate the uncertainties of our methods, including the unfolding procedure, and show that the overall accuracy exceeds that of the method used in the original studies of the KASCADE experiment.}

\maketitle

\newpage

\section{Introduction}
\label{sec:intro}
Mass composition is one of the main problems in the physics of high-energy cosmic rays (CR)~\cite{Gabici:2019jvz}.
Despite decades of experimental studies the precise mass composition above the so-called knee of the CR spectrum ($E \gtrsim 10^{15}$~eV) is not known. While the general trend of composition becoming heavier with energy is expected from theory~\cite{1961NCim...22..800P, Gaisser:2011klf}, the results of the various cosmic rays experiments are incompatible with each other~\cite{Apel:2013uni, IceCube:2019hmk, TelescopeArray:2020bfv}. The knowledge of the spectra of separate mass components in the $10^{15} < E < 10^{18}$~eV energy range is important for the understanding of their origin, in particular, the transition between the galactic and the extragalactic cosmic rays is expected somewhere in this range~\cite{Gabici:2019jvz}.

The analysis of CR properties in this energy range is complicated by several issues. Unlike CRs of lower energies, these particles can be detected only indirectly, via so-called extensive air showers (EAS) of secondary particles, that they produce in the Earth's atmosphere. The longitudinal and lateral structure of these showers can be recorded by various experimental techniques: Cherenkov and fluorescence light detection~\cite{TelescopeArray:2020bfv}, detection of charged particles shower on the Earth surface and muon part of the shower underground~\cite{Apel:2013uni, IceCube:2019hmk}. Then the properties of the primary particle can be reconstructed using this data. It is relatively simple to reconstruct the direction of the primary CR from the geometry of the shower and the CR energy from both the geometry and lateral distribution function (LDF) of secondary particles. At the same time, the accurate reconstruction of the primary particle type is much more difficult since the showers produced by different nuclei are not that different. One can either look for a longitudinal development of the shower in the atmosphere deriving the atmospheric depth of the maximum light emitted by the shower, the so-called depth of shower maximum ($X_{\rm max}$), that is a proxy of a primary particle mass. Alternatively, one can analyze the charged particle content of the shower on the ground and its muonic content underground --- the heavier the primary nucleus --- the larger the total number of muons. Then, comparing these observables with the Monte-Carlo simulations of air showers from various primary nuclei one can hope to reconstruct the mass composition of primary CRs or even the energy spectra of separate mass components of the CR flux. The problem here is that all the mentioned observables are dependent on the hadronic interaction models used in the Monte-Carlo simulations. So the mass component spectra reconstructed from one and the same experimental data using different Monte-Carlo models may differ significantly~\cite{kascade_cuts}.

In the present study, we aim to develop modern methods to reanalyze the original data of the KASCADE experiment and to re-derive the CR mass components spectra from it.
The method we choose for this analysis is machine learning (ML). Its benefits for the field of cosmic-ray experiments data analysis were proven by many recent studies~\cite{IceCube:2019hmk, Ivanov:2020nfo, Kalashev:2021vop, TelescopeArray:2018bep, Erdmann:2017str, PierreAuger:2021fkf}. The key idea behind the applicability of these methods is that the data from surface detectors about an EAS event can effectively be interpreted as an image, which opens the way for extensive usage of the machine learning methods developed for image analysis recently. Another idea is that the information viable for mass composition analysis of CRs is likely not accumulated in a few observables of EAS by rather dispersed in the whole ``image'' of the EAS footprint, so that the ML methods that are working with a huge number of variables is expected to efficiently ``convert'' this image back to a type of the primary particle.   

The paper is organized as follows. First, we introduce the KASCADE experiment, its data, Monte-Carlo, and reconstruction in Section~\ref{sec:exp-MC}. Then we present the machine learning methods we develop for this study in Section~\ref{sec:ML}. We make an initial estimation of the performance of all these ML methods using KASCADE Monte-Carlo in Section~\ref{sec:perf}. In Section~\ref{sec:tests} we perform several tests and estimate the main uncertainties for the convolutional neural network method (CNN) that showed the best performance. In Section~\ref{sec:unfold} we discuss and test the unfolding procedure for the reconstruction of the separate mass components spectra. We estimate all uncertainties related to this procedure, as well as the overall accuracy of the mass components spectra reconstruction. We apply the CNN + unfolding to a small part of the KASCADE data set that we call ``unblind data'', and compare the results with the original KASCADE mass composition analysis. We conclude in Section~\ref{sec:conclusion}.

\section{Experiment, data and Monte-Carlo}
\label{sec:exp-MC}
KASCADE is an extensive air shower experiment that was located in KIT Campus North, Karlsruhe, Germany ($49.10^\circ$~N, $8.44^\circ$~E), at 110~m a.s.l, corresponding to an average atmospheric depth of $1022\; {\rm g/cm}^2$~\cite{KASCADE:2003swk}. It started to operate in 1996, underwent an extension called KASCADE-Grande in 2003, and finished the data acquisition in 2013. The KASCADE experiment (without Grande extension), which data is used in the present study, was comprised of 252 scintillator detectors placed in a rectangular grid with 13~m spacing, covering the area of $200 \times 200\; {\rm m}^2$ in total. This array is capable of detecting a secondary particle footprint of extensive air shower initiated by cosmic rays with primary energies in $\sim 500\; {\rm TeV}$ --- $100\; {\rm PeV}$ energy range. Detectors in an outer part of the grid contained a metal shielding layer between scintillator layers so that the upper scintillator layer detects a $e/\gamma$\nobreakdash-dominated signal while the lower one detects a muon-dominated signal. The KASCADE experiment also contained several other parts: a central detector, an underground muon tracking detector, etc., but in the present study, we only use the data from the main detector array.
A detailed description of the experiment is given in Refs.~\hbox{\cite{KASCADE:2003swk, KCDC_manual}}.

In the present study, we use the KASCADE Monte-Carlo sets and preselection data sets provided by the KASCADE Cosmic Ray Data Centre (KCDC)~\cite{Haungs:2018xpw}. The cosmic ray events in the datasets are reconstructed from the raw detector readings by the iterative algorithm using the Kascade Reconstruction for ExTensive Airshowers program (KRETA). 
The reconstruction gives the following parameters (features) for each cosmic ray event: shower core position ($x$, $y$), zenith angle ($\theta$), azimuth angle ($\phi$), number of electrons ($N_e$), number of muons ($N_\mu$) and shower age parameter ($s$). The parameters $N_e$, $N_\mu$ and $s$ are determined from the fit of the modified NKG LDF function~\cite{KCDC_manual}:
\begin{equation}
\rho_{e, \mu} = C(s) \cdot N_{e, \mu} \cdot \left(\frac{r}{r^{e, \mu}_m} \right)^{s-\alpha} \cdot \left(1+\frac{r}{r^{e, \mu}_m} \right)^{s-\beta};
\end{equation}
\begin{equation*}
C(s) = \frac{\Gamma(\beta-s)}{2\pi (r^{e, \mu}_m)^2 \Gamma(s-\alpha+2)(\alpha+\beta-2s)};
\end{equation*}
where $r$ is a distance to a shower core, $r^e_m = 89$~m and $r^\mu_m = 420$~m are Moliere radii for electrons and muons respectively, $\alpha$ and $\beta$ are numerical parameters determined from the Monte-Carlo simulations. 

\begin{figure}[t]
    \centering
    \includegraphics[width=0.6\linewidth]{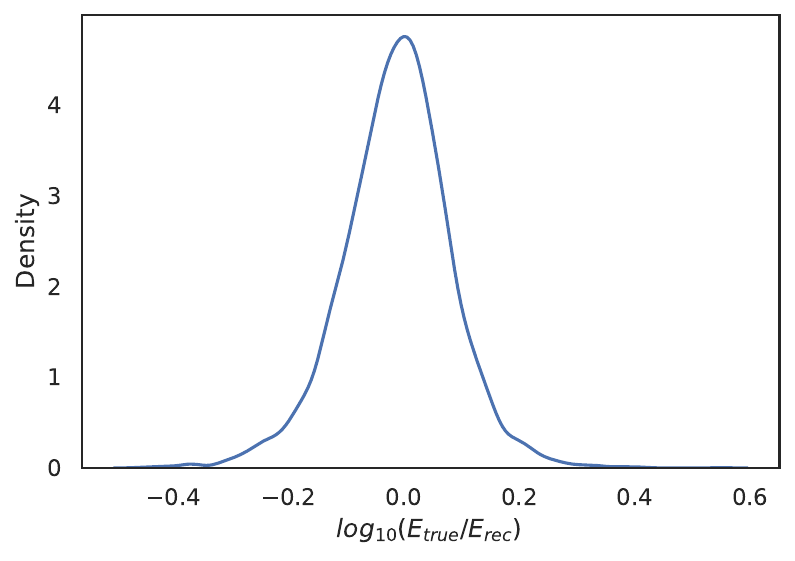}
    \caption{The distribution of the energy reconstruction uncertainty $\log_{10}{(E_{true}/E_{rec})}$ for the KASCADE standard method Eq.~\ref{eq:e_rec}.
    }
    \label{fig:default_energy_resolution_KASCADE}
\end{figure}

\begin{figure}[t]
    \centering
    \includegraphics[width=\linewidth]{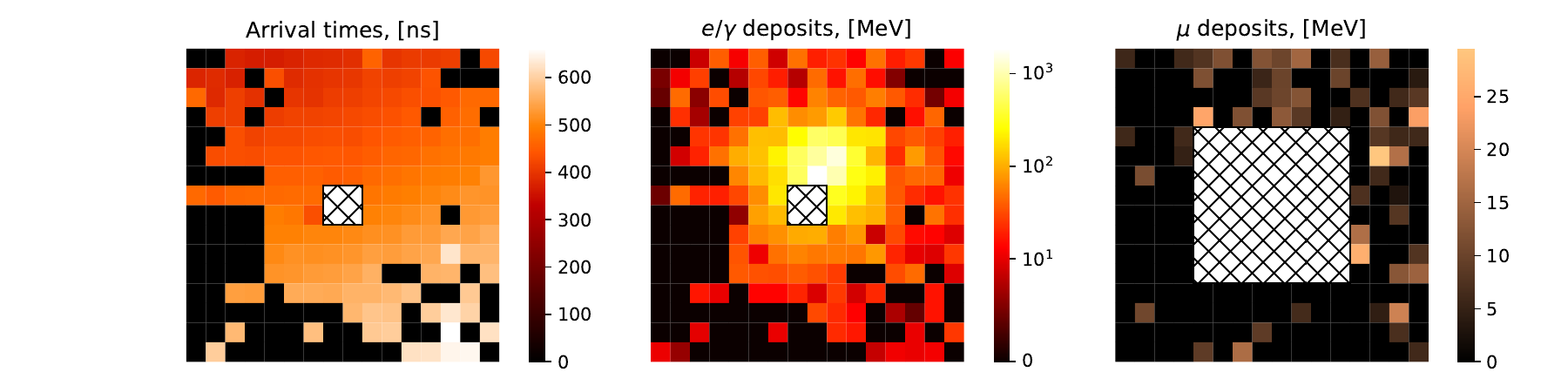}
    \caption{Example of the real event in the dataset. The matrices of arrival times, $e/\gamma$, and $\mu$ deposits are shown. Reconstructed features of the event are: $\log_{10} (E/\text{eV}) = 15.45$, $\theta = 19.37^{\circ}$. Note, that KASCADE does not have detector stations in the central 2x2 part for arrival times and $e/\gamma$ deposits, and in the central 8x8 part for $\mu$ deposits. These areas are represented in the figure with a strikethrough area.}
    \label{fig:exp_event_example}
\end{figure}

The value of the event energy $E$ is reconstructed from the parameters $N_e$ and $N_\mu$ with the following relation~\cite{Haungs:2018xpw}:
\begin{multline}
    \log_{10}(E/{\rm eV}) = 1.93499 + 0.25788\log_{10} (\hat{N_e}) + 0.66704\log_{10}(\hat{N_\mu}) +\\
    0.07507\log_{10}^2(\hat{N_e}) + 0.09277\log_{10}^2(\hat{N_\mu}) - 0.16131\log_{10}(\hat{N_e})\log_{10}(\hat{N_\mu})
    \label{eq:e_rec}
\end{multline}
where $\hat{N_e}$ and $\hat{N_\mu}$ are obtained from $N_e$ and $N_\mu$ by $\theta$-dependent rescaling to account for atmospheric attenuation, see Ref.~\cite{KCDC_manual} for details.
It is important to note that this formula is built using Monte-Carlo events based on high-energy hadronic interaction model QGSJet-II.02~\cite{Ostapchenko:2004ss} and low-energy hadronic interaction model FLUKA-2002.4~\cite{Ferrari:2005zk}. We computed the energy resolution and bias to be $9.5\%$ and $-1\%$ respectively, in terms of the logarithm of the simulated to the reconstructed energies ratio, $\ln (E_{\rm MC}/E_{\rm rec})$, for given quality cuts, additional energy cut and composition of the full Monte-Carlo set (see below). The full distribution of the uncertainty of the energy reconstruction is shown in Fig.~\ref{fig:default_energy_resolution_KASCADE}.
In addition to observables attributed to a CR event as a whole, there are values of $e/\gamma$ and $\mu$ energy deposits for separate detectors within the event. There are also arrival timings of the shower front for each detector but we do not use them in the present analysis. A graphical example of a recorded event is given in Fig.~\ref{fig:exp_event_example}.

The efficiency of event trigger and reconstruction depends on the type of primary particle, the full efficiency is reached at $E \gtrsim 10^{15}$~eV~\cite{KCDC_manual} for all primaries.
We use the quality cuts recommended by KASCADE~\cite{kascade_cuts}: ${\theta < 18^\circ}$, ${x^2 + y^2 < 91}$~m, ${\log_{10} N_e > 4.8}$, ${\log_{10} N_\mu > 3.6}$, and the cut on the shower age set by KCDC: ${0.2 < s < 1.48}$~\cite{KCDC_manual}, that is tighter than the original KASCADE cut (${0.2 < s < 2.1}$). The efficiency of these cuts also depends on a primary type, it reaches 99\% at $E > 10^{15.15}$~eV for protons and at $E > 10^{15.33}$~eV for iron nuclei.
We also introduce an additional cut on the event energy ${\log_{10}(E/{\rm eV})> 15.15}$. It is based on our study of confusion matrix and unfolding procedure stability.
This additional cut is applied to the real data sets and to Monte-Carlo test sets, but not to Monte-Carlo training sets (see discussion in a Sec.~\ref{sec:perf}). The total number of events in the data set after all cuts is $\sim 3.5 \cdot 10^6$. 
Also, the set with a looser zenith angle cut ${\theta < 30^\circ}$ is studied separately, it contains $\sim 8.6 \cdot 10^6$ events in total. In this zenith angle range, the experiment reaches full efficiency of events detection at $E > 10^{15.35}$~eV.
We divide the experimental data set into an unblind part containing 20\% of randomly selected events and a blind part containing the remaining 80\% of events. 
The unblind part is used for the consistency tests of the mass composition reconstruction methods developed in this study and for the estimation of their realistic uncertainties. The blind part of the data is not analyzed in this paper and is left for the full-scale mass composition study.

The KCDC service provides CORSIKA~\cite{heck1998corsika} simulations with events generated for five individual mass groups: \emph{p}, \emph{He}, \emph{C}, \emph{Si} and \emph{Fe} in energy range $10^{14} \le E \le 3 \cdot 10^{18}$~eV and zenith angle range $0 \le \theta \le 42^\circ$. Similar sets were produced for several hadronic interaction models, namely: QGSJetII-02~\cite{Ostapchenko:2004ss} + FLUKA~\cite{Ferrari:2005zk} (ver. 2002.4), that were used in the latest original KASCADE mass composition analysis~\cite{Apel:2013uni} ($\sim 3.7 \cdot 10^4$ events after quality cuts and additional energy cut); QGSJetII-04~\cite{Ostapchenko:2010vb} + FLUKA (ver. 2012.2.14\_32) ($\sim 1.3 \cdot 10^5$ events); EPOS-LHC~\cite{Pierog:2013ria} + FLUKA (ver. 2011.2b.4\_32) ($\sim 6.7 \cdot 10^4$ events) and Sibyll\,2.3c~\cite{Riehn:2015aqb} + FLUKA (ver. 2011.2c.3\_64) ($\sim 6.8 \cdot 10^4$ events).
As the available amount of Monte-Carlo events is much less than the amount of data events we are taking into account statistical uncertainties of the Monte-Carlo in our results. 
For each hadronic model, the number of protons and light nuclei events in the MC sets exceeds the number of heavy nuclei but the ratio is smaller than 2. For instance for the set of QGSJetII-04 hadronic model the proportion for \emph{p}:\emph{He}:\emph{C}:\emph{Si}:\emph{Fe} is 0.28:0.22:0.19:0.16:0.15. For other hadronic models, the numbers are quite similar. The energy spectrum was set to $E^{-2}$ in CORSIKA simulations and corrected to $E^{-2.7}$ in later stages of analysis. The resulting Monte-Carlo sets were reconstructed using the same code and the actual detector response as for the real data and contained the same observables.

\section{Machine learning methods}
\label{sec:ML}
In the paper, we use a set of different machine learning (ML) methods for event-by-event mass group classification.
We are starting with Random Forest~(RF) --- the classical ML approach, which is our baseline. It takes as input only the reconstructed event features, but not the detectors deposits.
Then we switch to using neural networks (NN) to incorporate more data in the analysis.
All NN models take as input both event features and deposits from $e/\gamma$ and $\mu$ detectors.
In particular, we build a convolutional neural network~(CNN), a simple multi-layer perceptron~(MLP), and an EfficientNet.
All the models were trained with three MC datasets corresponding to three hadronic interaction models: QGSJet-II.04, EPOS-LHC and Sibyll\,2.3c separately. The set of the QGSJet-II.02 model is also used separately for direct comparison with the original KASCADE mass composition analysis.
These datasets were divided into train, validation, and test subsets.
Throughout the study, we use train and validation subsets for training models (the validation set is used for early stopping to avoid overfitting and hyperparameter tuning of the models).
All the metrics are evaluated on test subsets of the corresponding MC datasets.

\subsection{Random Forest}
\label{sec:ML:RF}
It was chosen to use Random Forest \cite{ho1995random} as a baseline solution for the classification of individual events for five mass groups.
The following reconstructed features of the air shower were used as input parameters: energy, shower core coordinates, arrival direction, muon and electron numbers, and shower age.
We trained two models: classifier and regressor.
The classifier predicts the particle type directly while the regressor predicts the mass of the particle, from which we determine the particle type.
The hyperparameters of both models were optimized using the Grid Search algorithm, the details are given in Appendix~\ref{app:arch}.
Both RF models were implemented and trained using scikit-learn package~\cite{scikit-learn}.

\subsection{Multi-layer perceptron}
\label{sec:ML:MLP}
The simplest neural network we use is a feed-forward multilayer perceptron consisting of two hidden layers, each followed by batch normalization, ELU activation, and dropout (with a rate of 0.15).
We use integral signals from the $e/\gamma$ and $\mu$ detector stations and reconstructed zenith and azimuth angles as input for the model.
The model was trained with the Adam algorithm~\cite{kingma2014adam} using a batch size of 1024 and a variable learning rate (starting from 0.001, then multiplying it by 0.5 when validation loss plateaus) until validation loss stops improving.
This quite primitive architecture shows quite good performance, nevertheless (see next Section).
Detailed architecture of the model is given in Appendix~\ref{app:arch}.

\subsection{Convolutional Neural Network}
\label{sec:ML:CNN}
The second NN we use is a convolutional neural network.
This approach allows us to include the deposits from the detectors in the analysis. 
We developed and trained a simple convolutional neural network, inspired by the LeNet-5~\cite{lenet_arch} model, to classify individual events for five mass groups. 
The integral signals from the $e/\gamma$ and $\mu$ detector stations for each event are fed into the model. Additionally, we append the high-level features $\log_{10} N_e$, $\log_{10} N_{\mu}$,~$\theta$,~$s$ as an input to the first dense layer. We need to note that the CNN model is trained with Monte-Carlo sets with the looser zenith angle cut: $\theta < 30^\circ$. The details of this ML model can be found in Appendix~\ref{app:arch}.

\subsection{EfficientNet}
\label{sec:ML:effN}
As an additional benchmark, we trained an EfficientNetV2-S~\cite{tan2021efficientnetv2} model that belongs to a family of new, high-efficiency convolutional neural networks.
EfficientNetV2 uses some specific hardware optimizations, such as the use of Fused-MBConv convolutions instead of depthwise convolutions in early layers, that are combined with the heavy use of neural architecture search.
The model takes integral signals from the $e/\gamma$ and $\mu$ detector stations as a main input for convolutional layers, and reconstructed zenith and azimuth angles as an additional input for the bottleneck layer of the model.
The model was trained with the Adam algorithm using a batch size of 128 and a variable learning rate (starting from 0.001, then multiplying it by 0.5 when validation loss plateaus) until validation loss stops improving.
While EfficientNetV2-S is considerably larger than our CNN, it achieves almost identical performance, which suggests that more complex models may not lead to significant improvements in performance when dealing with this particular problem. Furthermore, the larger model size could lead to increased computational costs and longer training times, making it less practical for our purposes. Therefore, we do not discuss the results of this ML model further.

\section{Performance of the methods}
\label{sec:perf}
In this section, we present the basic results of cosmic ray events primary mass classification using our machine learning methods. 
We compare their quality and reconstruct Monte-Carlo spectra to check for correctness.
All results here and further are computed on the Monte-Carlo test sets or the unblind data set using the quality cuts described in the Section~\ref{sec:exp-MC} plus the additional cut $\log_{10} (E/\text{eV}) > 15.15$ that ensures the full efficiency of detection for all primaries. However, we do not use this cut for the MC training sets: in general for machine learning the total amount of training data is crucial, while their possible non-uniformities are of secondary importance.

\subsection{Confusion matrices}
\begin{figure}[t]
    \captionsetup[subfigure]{aboveskip=0pt,belowskip=-7pt}
    \centering
    \begin{subfigure}{.32\textwidth}
        \includegraphics[width=\textwidth]{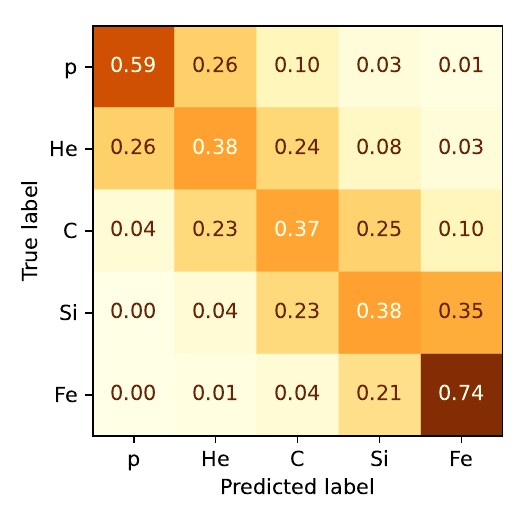}
       \caption{}
    \end{subfigure}
    \hfill
    \begin{subfigure}{.32\textwidth}
        \includegraphics[width=\textwidth]{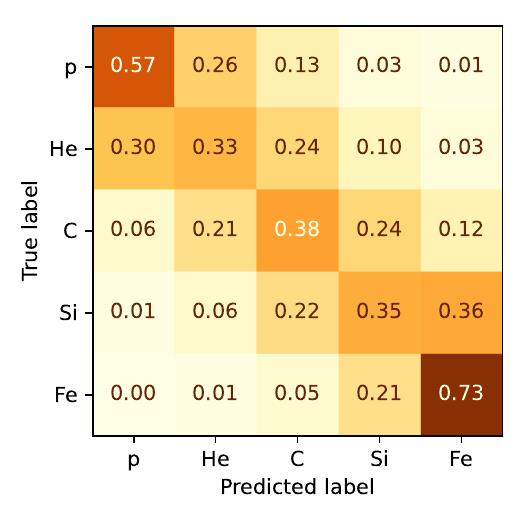}
       \caption{}
    \end{subfigure}
    \hfill
    \begin{subfigure}{.32\textwidth}
        \includegraphics[width=\textwidth]{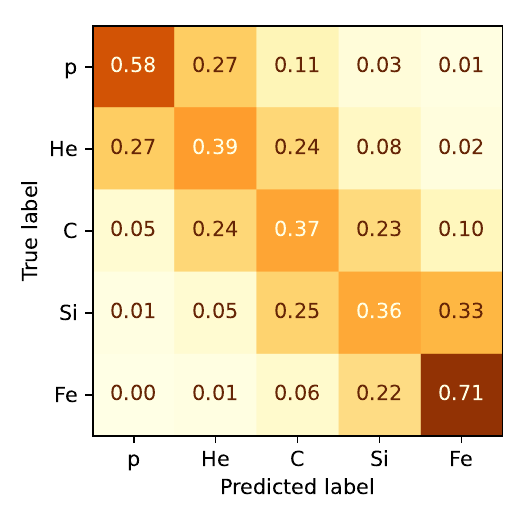}
       \caption{}
    \end{subfigure}
     
    \caption{Confusion matrices for RF Classifier models trained and tested with three different hadronic interaction models (a) QGSJet-II.04, (b) EPOS-LHC and (c) Sibyll\,2.3c.}
    \label{rf_confusion_matrices}
\end{figure}
\begin{figure}[t]
    \captionsetup[subfigure]{aboveskip=0pt,belowskip=-7pt}
    \centering
    \begin{subfigure}{.32\textwidth}
        \includegraphics[width=\textwidth]{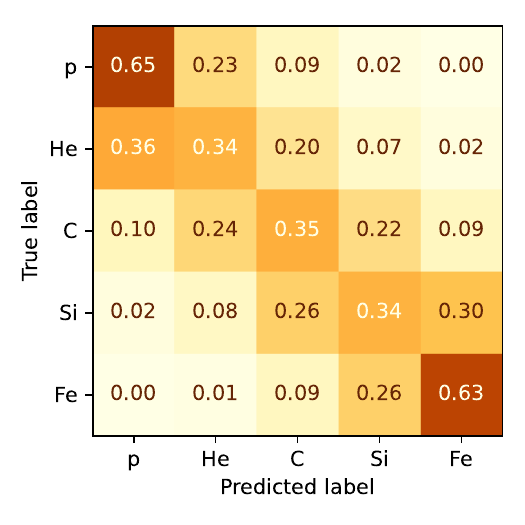}
       \caption{}
    \end{subfigure}
    \hfill
    \begin{subfigure}{.32\textwidth}
        \includegraphics[width=\textwidth]{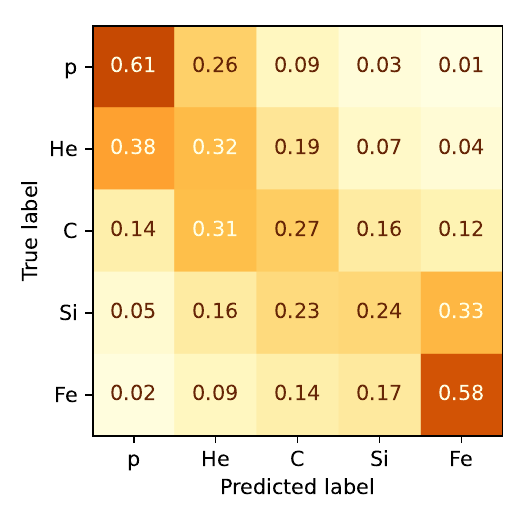}
       \caption{}
    \end{subfigure}
    \hfill
    \begin{subfigure}{.32\textwidth}
        \includegraphics[width=\textwidth]{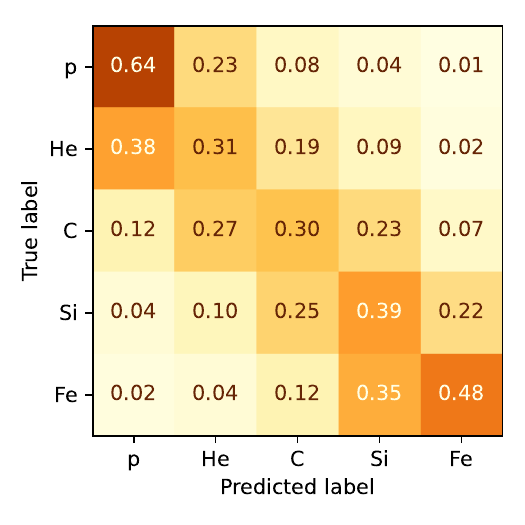}
       \caption{}
    \end{subfigure}
    \caption{Confusion matrices for MLP models trained and tested with three different hadronic interaction models (a) QGSJet-II.04, (b) EPOS-LHC and (c) Sibyll\,2.3c.}
    \label{mlp_confusion_matrices}
\end{figure}
\begin{figure}[t]
    \captionsetup[subfigure]{aboveskip=0pt,belowskip=-7pt}
    \centering
    \begin{subfigure}{.32\textwidth}
        \includegraphics[width=\textwidth]{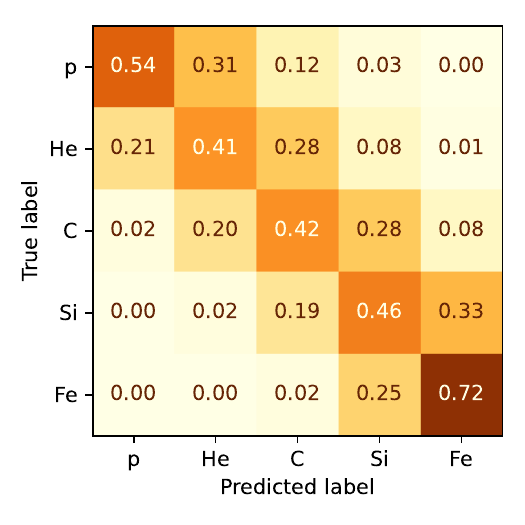}
       \caption{}
    \end{subfigure}
    \hfill
    \begin{subfigure}{.32\textwidth}
        \includegraphics[width=\textwidth]{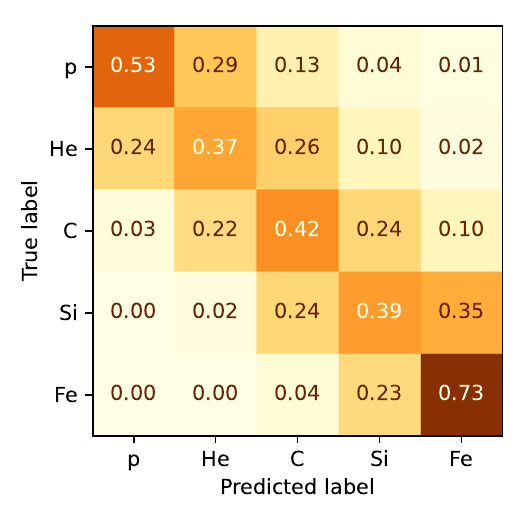}
       \caption{}
    \end{subfigure}
    \hfill
    \begin{subfigure}{.32\textwidth}
        \includegraphics[width=\textwidth]{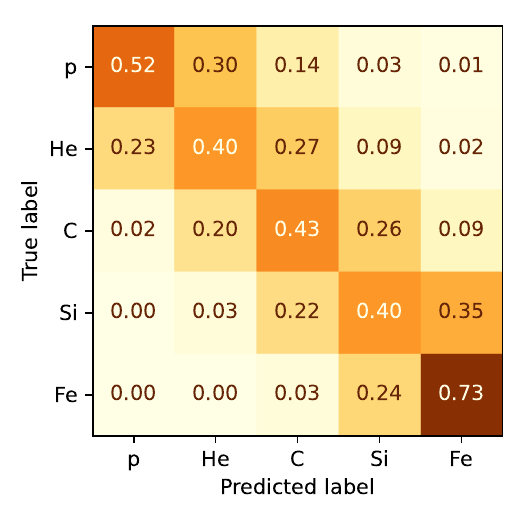}
       \caption{}
    \end{subfigure}
    \caption{Confusion matrices for CNN models trained and tested with three different hadronic interaction models (a) QGSJet-II.04, (b) EPOS-LHC and (c) Sibyll\,2.3c.}
    \label{cnn_confusion_matrices}
\end{figure}

\label{sec:perf:conf-mat}
The simplest way to estimate the quality of the trained ML models is to compute confusion matrices on the test datasets for each model.
The confusion matrix shows the proportion of the particle of each type that the given model classifies as this type and all other types. A more diagonal matrix means more precise classification. Also, the transposed confusion matrix is used as a response matrix for the unfolding procedure described later.
The confusion matrices for the ML models used in this paper are shown in Fig.~\mbox{\ref{rf_confusion_matrices}~-- \ref{cnn_confusion_matrices}}. One can see that the matrices are diagonal in most cases. The diagonal elements have much better accuracy than the random guessing (0.2 for 5 components classification). The best performance is shown by CNN. 

\subsection{Reconstruction of mass components spectra}
\label{sec:perf:spec-MC}
\begin{figure}[t]
     \includegraphics[width=\textwidth]{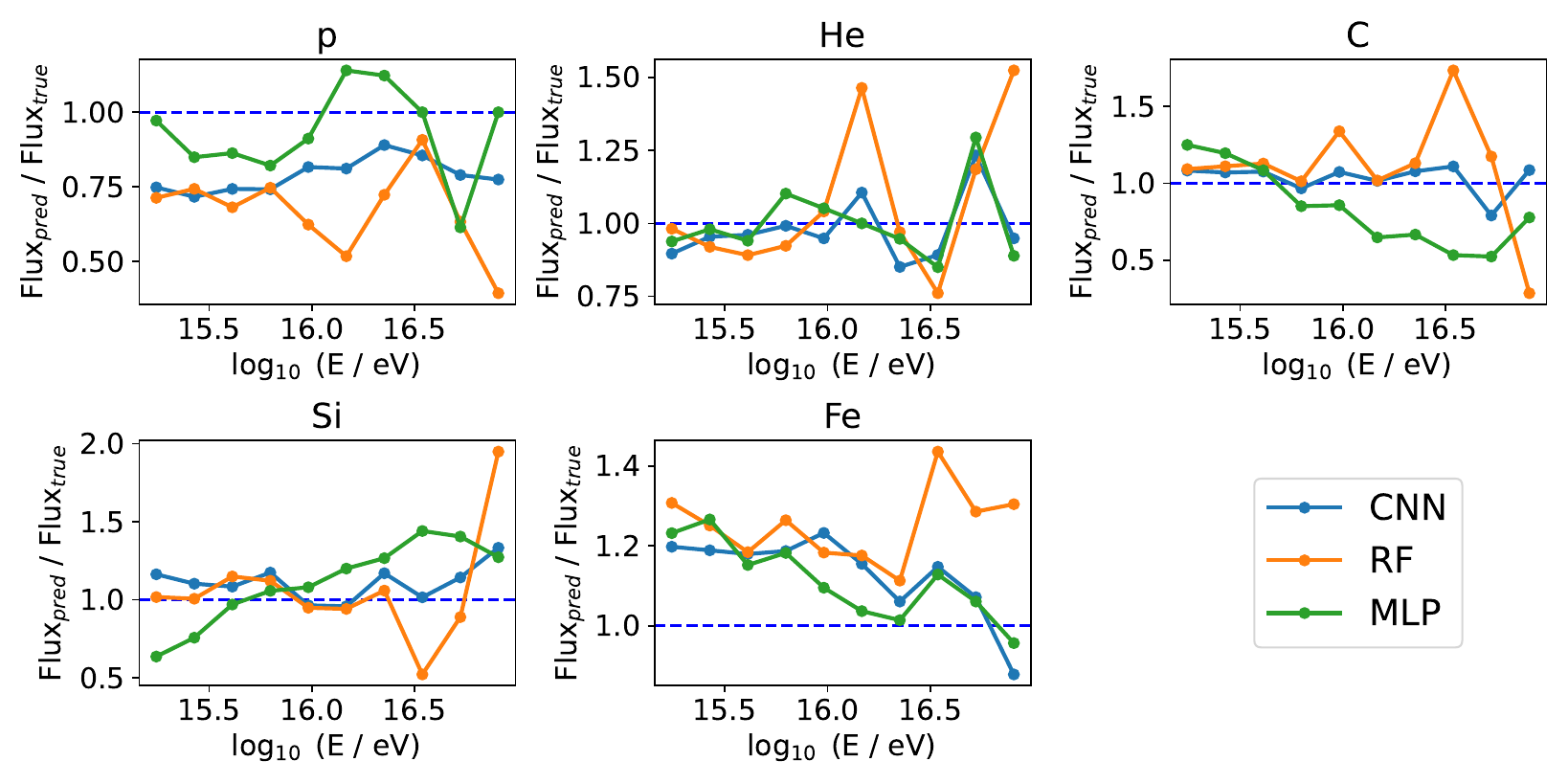}
     \caption{Results of the deviation of predicted spectra from true spectra in Monte-Carlo test sets~(QGSJet-II.04) for the CNN, RF and MLP models.
     }
    \label{fig:reconstructed mc spectra}
\end{figure}

In Fig.~\ref{fig:reconstructed mc spectra} we show the discrepancy between the true spectra and the reconstructed spectra of separate mass components for the QGSJet-II.04 model and our reconstructions with RF, CNN, and MLP. 
One can see that while all ML models have roughly similar accuracy in energy spectra reconstruction in general, the accuracy of the CNN model is the least energy-dependent. The general underestimation of proton flux and overestimation of iron flux are subject to further correction with the unfolding procedure (see Section~\ref{sec:unfold}).

\subsection{Mass components reconstruction in mixtures}
\label{sec:tests:mix}
\begin{figure}[t]
        \includegraphics[width=\linewidth]{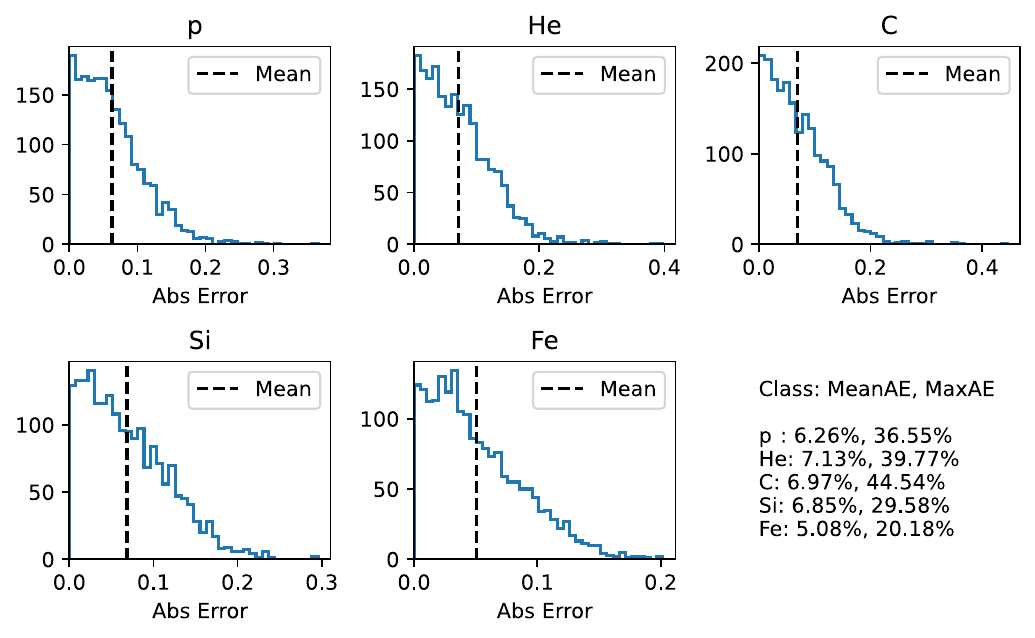}
        \includegraphics[width=\linewidth]{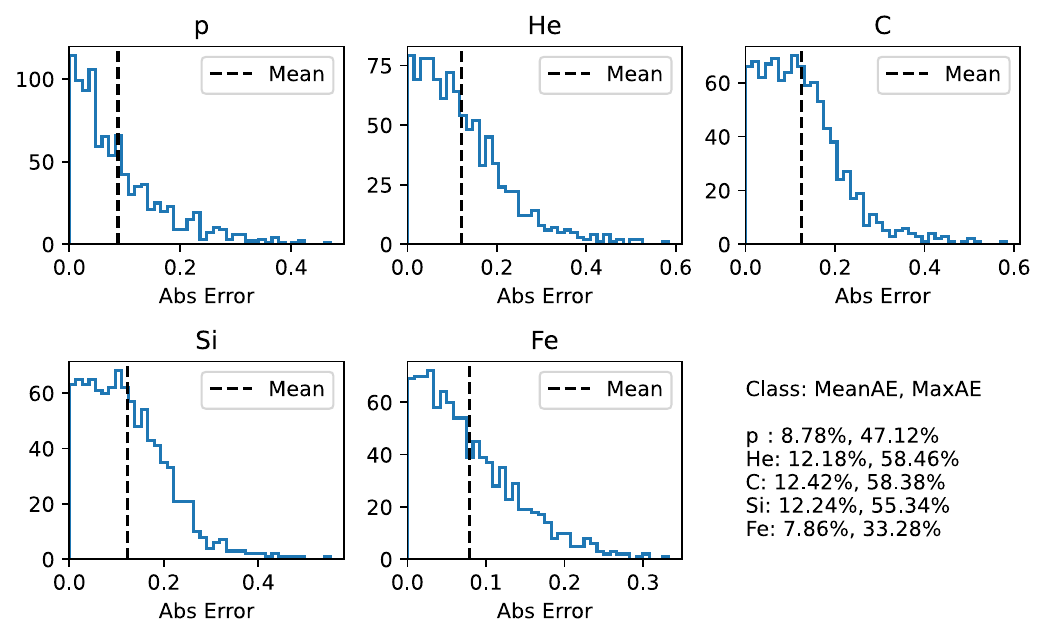}
    \caption{Distributions of the absolute errors between true and predicted fractions for the CNN model trained and tested with QGSJet-II.04 hadronic interaction model
    (upper figure)~in the ``uniform'' mixtures approach,
    (lower figure)~in the ``grid'' mixtures approach.
    Black lines in the figures represent mean absolute error~(MeanAE) for each mass component. MaxAE means maximum absolute error for each mass component.}
    \label{fig:mixtures systematics}
\end{figure}

Another way to estimate the quality of the model predictions is to calculate the mean absolute error (MAE) between true component fractions in ensembles and the predicted fractions~\cite{Kalashev:2021vop}. The smaller the MAE for a particular component, the better the ML model does its job. This method allows us to estimate how the accuracy of the reconstruction of particular mass components depends on the type of mixture under analysis.

To perform this estimation we need to create a set of various mass components mixtures. We produce 2\,000 random ensembles of 5\,000 events each using the test set. There are two obvious methods to make such an ensemble.
In the first method, the fractions of components in a given ensemble are distributed uniformly so we call this approach ``uniform''. 
It has the disadvantage that extreme mixtures (for example pure protons) are very rare, while the sets with approximately equal fractions are much more common.
Therefore, in addition, we use the second method of creating mixtures, where the proportions of mass components are distributed over a grid with a step of~0.1 from~0 to~1.
This approach is called ``grid''.
The distributions of the absolute errors between true and predicted fractions in these two approaches are shown in Fig.~\ref{fig:mixtures systematics}.
One can see that the accuracy of the ML reconstruction is much better for the uniform sets than the accuracy averaged over all possible sets generated by the ``grid'' approach. 

\begin{table}[t]
    \centering
    \begin{tabular}{|c|c|c|c|c|c|}
         \hline
         & \emph{p} & \emph{He} & \emph{C} & \emph{Si} & \emph{Fe} \\
         \hline
         RF & \textbf{0.083} & 0.123 & 0.127 & 0.123 & 0.085 \\
         \hline
         CNN & 0.088 & \textbf{0.122} & \textbf{0.124} & \textbf{0.122} & \textbf{0.079} \\
         \hline
         MLP & 0.089 & 0.128 & 0.136 & 0.131 & 0.089 \\
         \hline
    \end{tabular}
    \caption{Mean absolute error between true and reconstructed fractions of elements for three different models and mass groups calculated with ``grid'' approach using QGSJet-II.04 hadronic interaction model.
    The best result for each mass component is highlighted in bold.}
    \label{tab:CNN_MAE}
\end{table}
The ``grid'' approach covers more possible mixtures and gives a more strict estimate of the deviation. Therefore, we use this approach to compare our models.
The results for different models in the ``grid'' approach are shown in Table~\ref{tab:CNN_MAE}.
We see that all the models have a similar quality, but the CNN has slightly better performance.
The similarity in the quality of the three ML models, that were built using completely different ML techniques, may reflect the fact that we extracted the maximum available information about mass composition from the given input data. 

\section{Tests and uncertainties}
\label{sec:tests}
To ensure the correctness of the developed ML models we perform several consistency tests and estimate several uncertainties. Namely,
we study the behavior of the models for data with missed detector stations. We perform an ablation study to estimate the importance of various input parameters. We also estimate the dependence of the ML model performance on the zenith angle and the energy of the primary particle.
In this section, we consider the CNN model as our baseline model, while several tests performed for other ML models can be found in Appendix~\ref{app:rf-tests}.

\subsection{Ablation study}
\label{sec:tests:ablation}
To estimate the impact of the different input features on the ML method performance, we make an ablation study. 
We remove input features sequentially, train the model again, and compare the resulting confusion matrices.
This procedure helps us to understand the impact of different features on the result. It also tests that the model is focused on physical features and dependencies rather than simulation artifacts.

\begin{figure}[t]
    \centering
    \begin{subfigure}{.3\linewidth}
        \includegraphics[width=\linewidth]{figs/cnn/final_redraw/cm_qgs.pdf}
        \caption{}
    \end{subfigure}
    \hfill
    \begin{subfigure}{.3\linewidth}
        \includegraphics[width=\linewidth]{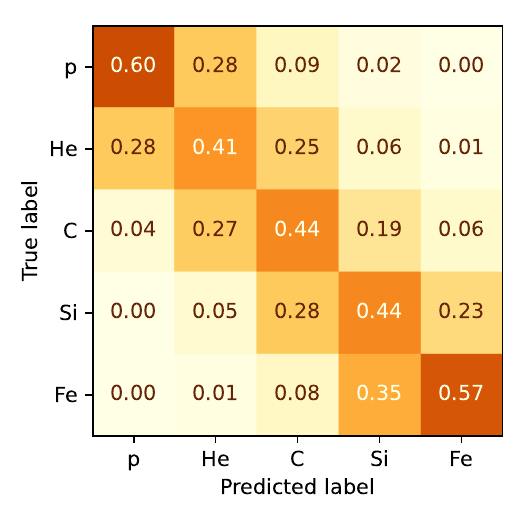}
       \caption{}
    \end{subfigure}
    \hfill
    \begin{subfigure}{.3\linewidth}
        \includegraphics[width=\linewidth]{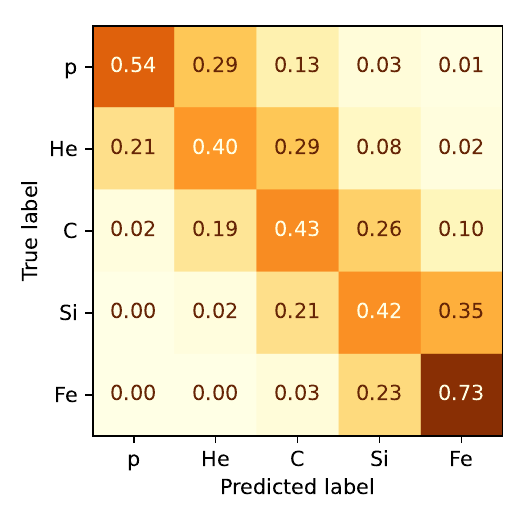}
       \caption{}
    \end{subfigure}
    \caption{Ablation study: confusion matrices (a)~of the default CNN, (b)~of the CNN trained only with $e/\gamma$ and $\mu$ detectors deposits, (c)~of the CNN trained only with reconstructed features~($\log_{10} N_e$, $\log_{10} N_\mu$, $s$, $\theta$).
    QGSJet-II.04 hadronic model is used for both training and test datasets.
}
    \label{fig:ablation study cnn}
\end{figure}

The results of the ablation study are shown in Fig.~\ref{fig:ablation study cnn}. 
We compare the confusion matrices of the default CNN trained with detectors deposits and reconstructed features, the CNN trained with deposits only and the CNN trained with reconstructed features only.
One can see that the matrix for reconstructed features only looks slightly better than the matrix for deposits only. 
It was found that this difference comes mostly from the zenith angle~($\theta$)
In general, the other features complement each other so that the default CNN trained on deposits and reconstructed features shows slightly better performance with the most diagonal confusion matrix. 

\subsection{Uncertainty of missed detectors in data}
\label{sec:tests:miss-det}
\begin{figure}[t]
    \centering
    \includegraphics[width=\linewidth]{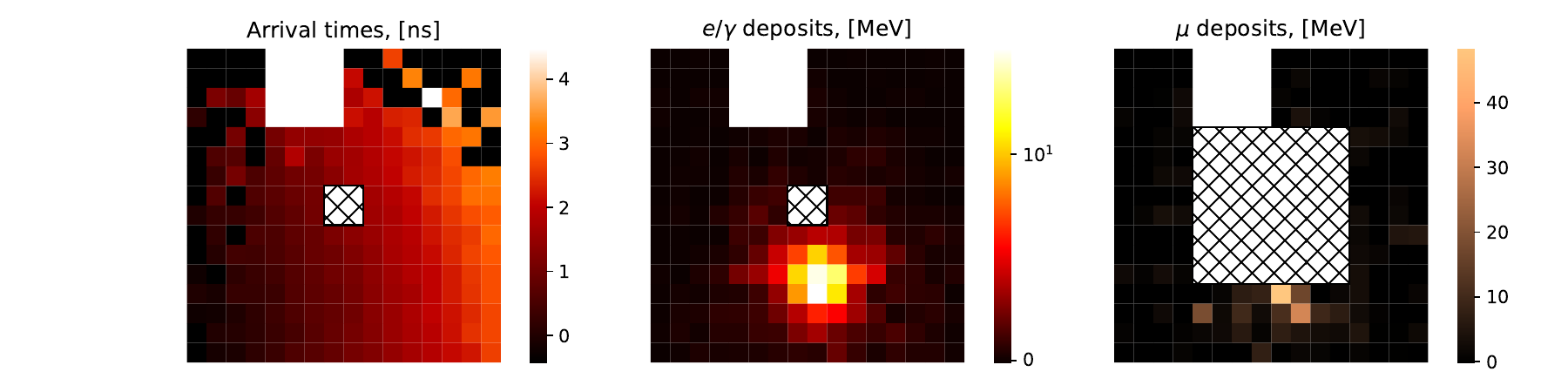}
    \caption{Example of the simulated event in the ``corrupted'' dataset. The white areas in the event represent missing detectors. The matrices of arrival times, $e/\gamma$, and $\mu$ deposits are shown. Reconstructed features of the event: $\log_{10} (E / \text{eV}) = 15.61$, $\theta = 15.68^{\circ}$.
    }
    \label{fig:missing detectors data representation}
\end{figure}

The experimental data contains events with non-working detector stations, while in KCDC Monte-Carlo this feature is not simulated. As our ML models are trained with the Monte-Carlo simulations, their performance with the data can be unstable due to this difference. 
We will consider the possible effect of this difference as a systematic error.
To estimate it, we prepared a ``corrupted'' data set from the test Monte-Carlo set, resembling the experimental data set with non-working stations. Each event in this set has a random region of $4 \times 4$ non-working stations.
An example of such an event is shown in Fig.~\ref{fig:missing detectors data representation}.

\begin{figure}[t]
    \centering
    \begin{subfigure}{.8\textwidth}
        \begin{subfigure}{.45\textwidth}
            \includegraphics[width=\textwidth]{figs/cnn/final_redraw/cnn_qgs_on_qgs_0reco.pdf}
            \caption{}
        \end{subfigure}
        \hfill
        \begin{subfigure}{.45\textwidth}
            \includegraphics[width=\textwidth]{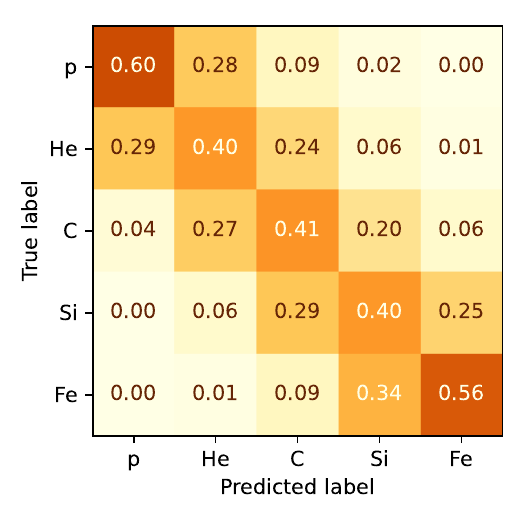}
           \caption{}
        \end{subfigure}
    \end{subfigure}
    \caption{Confusion matrices of the CNN: (a) trained with $e/\gamma$, $\mu$ deposits only,
    (b) the same but with ``corrupted'' data. Monte-Carlo of \mbox{QGSJet-II.04} hadronic interaction model is used for training and testing.
    }
    \label{fig:missing detectors cnn}
\end{figure}

In Fig.~\ref{fig:missing detectors cnn} we compare the confusion matrices of the CNN model obtained with the original test Monte-Carlo set and the ``corrupted'' Monte-Carlo set.
For this test, we use the CNN model trained with detector deposits only to estimate the impact of this uncertainty conservatively.
One can see that the general structure of the confusion matrix is conserved and the average difference of the diagonal values is about 2\%.
In what follows we consider this effect as a separate contribution to the total systematic uncertainty of our analysis method. As $\sim 50\%$ of events in the real data set contain non-working detectors, we take into account the estimated uncertainty with a factor of 0.5.

\subsection{Zenith angle dependence}
\label{sec:tests:zenith-dep}
\begin{figure}[t]
    \centering
    \begin{subfigure}{\linewidth}
        \includegraphics[width=\linewidth]{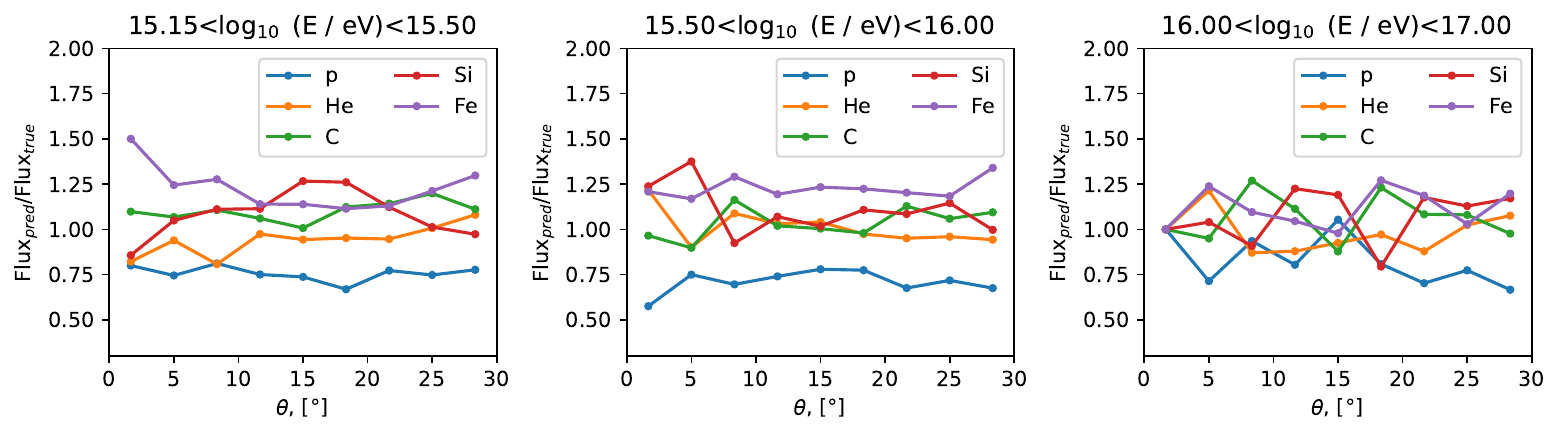}
        \caption{}
    \end{subfigure}
    \hfill
    \begin{subfigure}{\linewidth}
        \includegraphics[width=\linewidth]{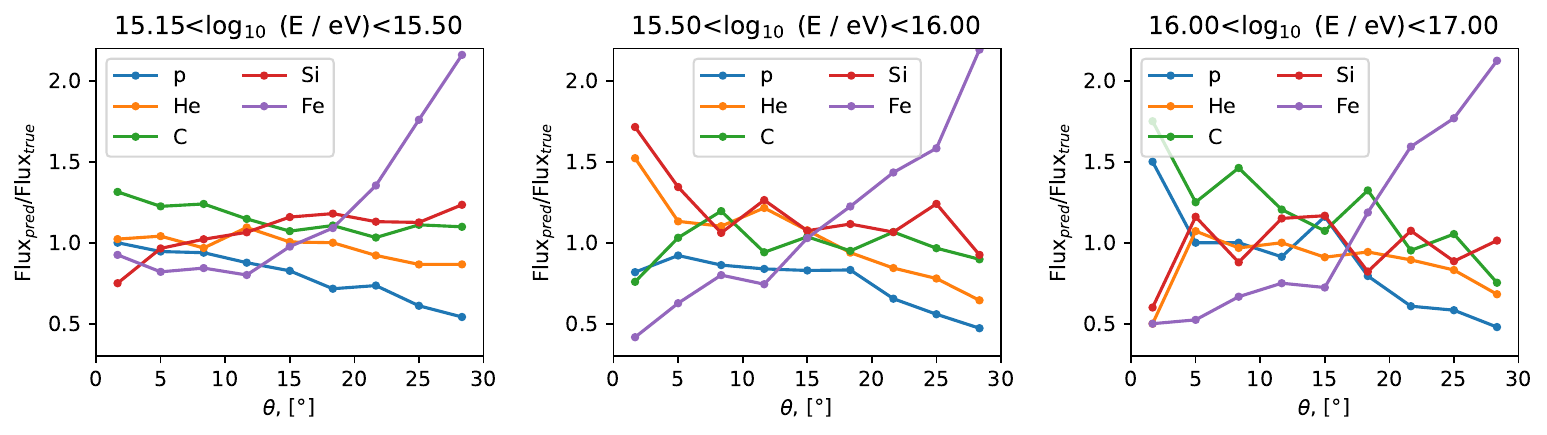}
       \caption{}
    \end{subfigure}
    \caption{
    The ratio of the predicted flux and the true flux as a function of the zenith angle for five mass groups in three different energy ranges. The figures are (a)~for the default CNN (that inputs $\theta$, $\log{N_e}$, $\log{N_\mu}$, $s$ and detector deposits), (b)~for the modified CNN without $\theta$ input parameter.}
    \label{fig:zenith angle dependence cnn}
\end{figure}

Ideally, the reconstructed mass component spectra should be independent of the zenith angle of the primary particle.
In Fig.~\ref{fig:zenith angle dependence cnn} we show the dependence of the inaccuracy of the reconstructed flux for separate mass components on the zenith angle in three different energy ranges.
We consider two variants of the CNN model: a default one and the one trained without zenith angle input. One can see that for CNN without $\theta$ input the error in \emph{Fe} component reconstruction grows dramatically with the angle. While there is no clear $\theta$-dependence in the results of the default CNN. We conclude that the addition of the zenith angle to the inputs ensures the stable behavior of the CNN model results.

\begin{figure}[t]
    \centering
    \begin{subfigure}{.8\textwidth}
        \begin{subfigure}{.45\textwidth}
            \includegraphics[width=\textwidth]{figs/cnn/final_redraw/cm_qgs.pdf}
            \caption{}
        \end{subfigure}
        \hfill
        \begin{subfigure}{.45\textwidth}
            \includegraphics[width=\textwidth]{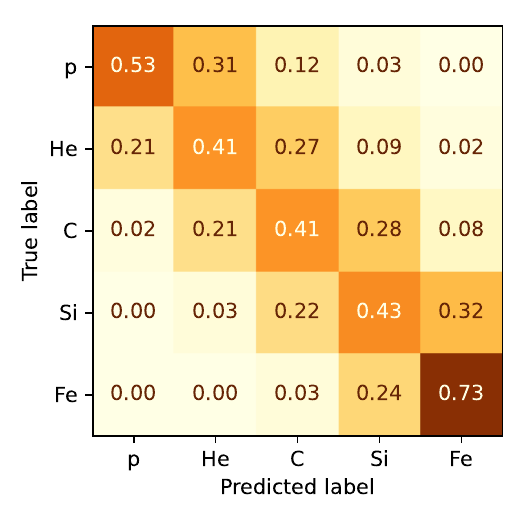}
           \caption{}
        \end{subfigure}
    \end{subfigure}
    
    \caption{Confusion matrices of the CNN with the different zenith angle cuts: (a)~$\theta < 18\degree$, (b)~$\theta < 30\degree$.
    CNN is trained and tested with QGSJet-II.04 hadronic model sets.
}
    \label{fig:different zenith angle cut cnn}
\end{figure}

\begin{figure}[t]
     \centering
    \begin{subfigure}{.48\textwidth}
        \includegraphics[width=\textwidth]{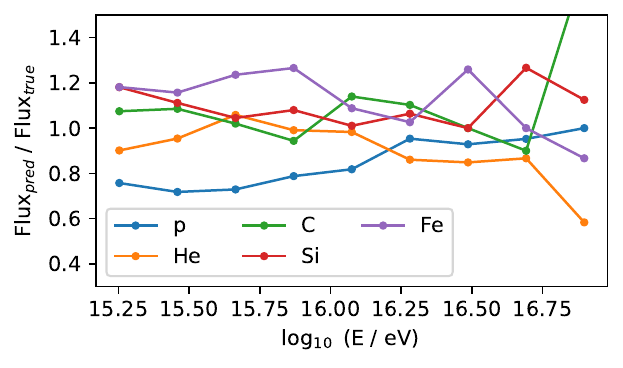}
        \caption{}
    \end{subfigure}
    \hfill
    \begin{subfigure}{.48\textwidth}
        \includegraphics[width=\textwidth]{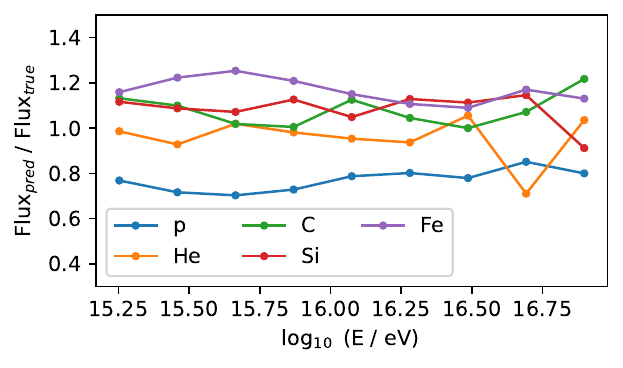}
       \caption{}
    \end{subfigure}
     \caption{The ratio of the predicted flux and the true flux as a function of energy for five mass groups for two zenith angle cuts:
     (a)~$\theta < 18\degree$ and (b)~$\theta < 30\degree$. CNN is trained and tested with QGSJet-II.04 hadronic model sets.
}
    \label{fig:different zenith angle cut cnn spectra}
\end{figure}

We also compare the confusion matrices for two zenith angle ranges (below 18$^\circ$ and below 30$^\circ$).
The results are shown in Fig.~\ref{fig:different zenith angle cut cnn}.
One can see that the matrices are almost similar, which shows us the validity of the ML model for a wide $\theta$ range.
The same conclusion can be derived from Fig.~\ref{fig:different zenith angle cut cnn spectra}, where we show the dependence of the reconstructed flux inaccuracy for separate mass components on the reconstructed energy in the same two zenith angle ranges.

\subsection{Energy dependence}
\label{sec:tests:energy-dep}
\begin{figure}[t]
    \centering
    \includegraphics[width=\linewidth]{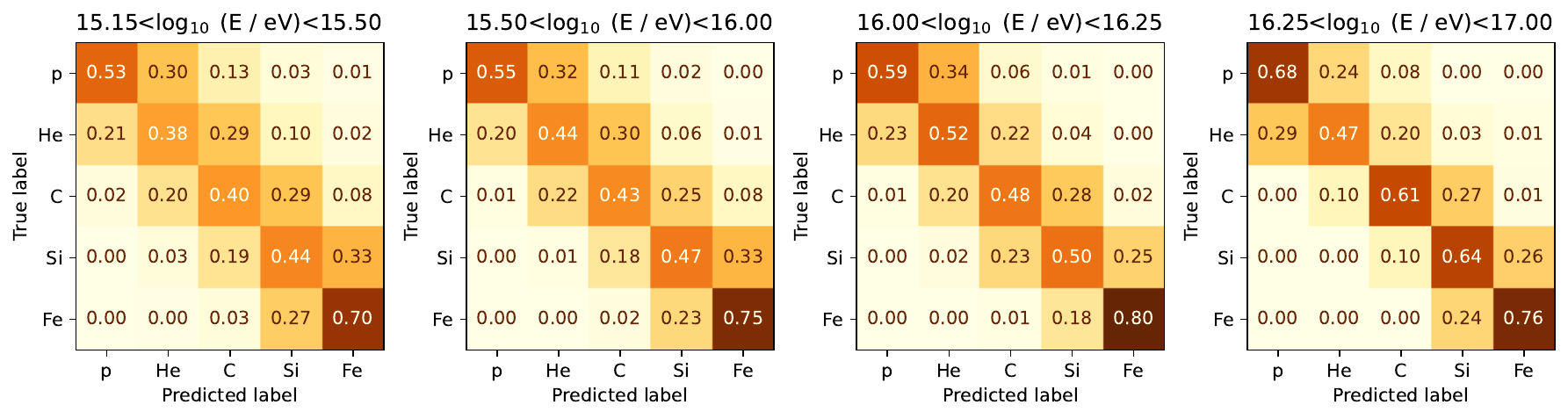}
    \caption{Confusion matrices of the CNN for different reconstructed energy bins. The neural network is trained and tested using the \mbox{QGSJet-II.04} hadronic interaction model.
    }
    \label{fig:confusion matrices energy bins}
\end{figure}

In this section, we compare the quality of the CNN predictions in the different energy bins. This is important for the unfolding procedure: if the predictions are stable in a certain energy range the response matrix can be averaged over this range to benefit from the larger statistics.
The CNN confusion matrices for different energy bins are shown in Fig.~\ref{fig:confusion matrices energy bins}. One can see that the accuracy of the CNN predictions is growing with energy. 
However we need to note that the behavior of the confusion matrix does not reflect the full picture: because of the steeply falling real spectrum, uncertainties at the higher energies are dominated by statistical fluctuations.

\begin{figure}[t]
    \centering
    \includegraphics[width=\linewidth]{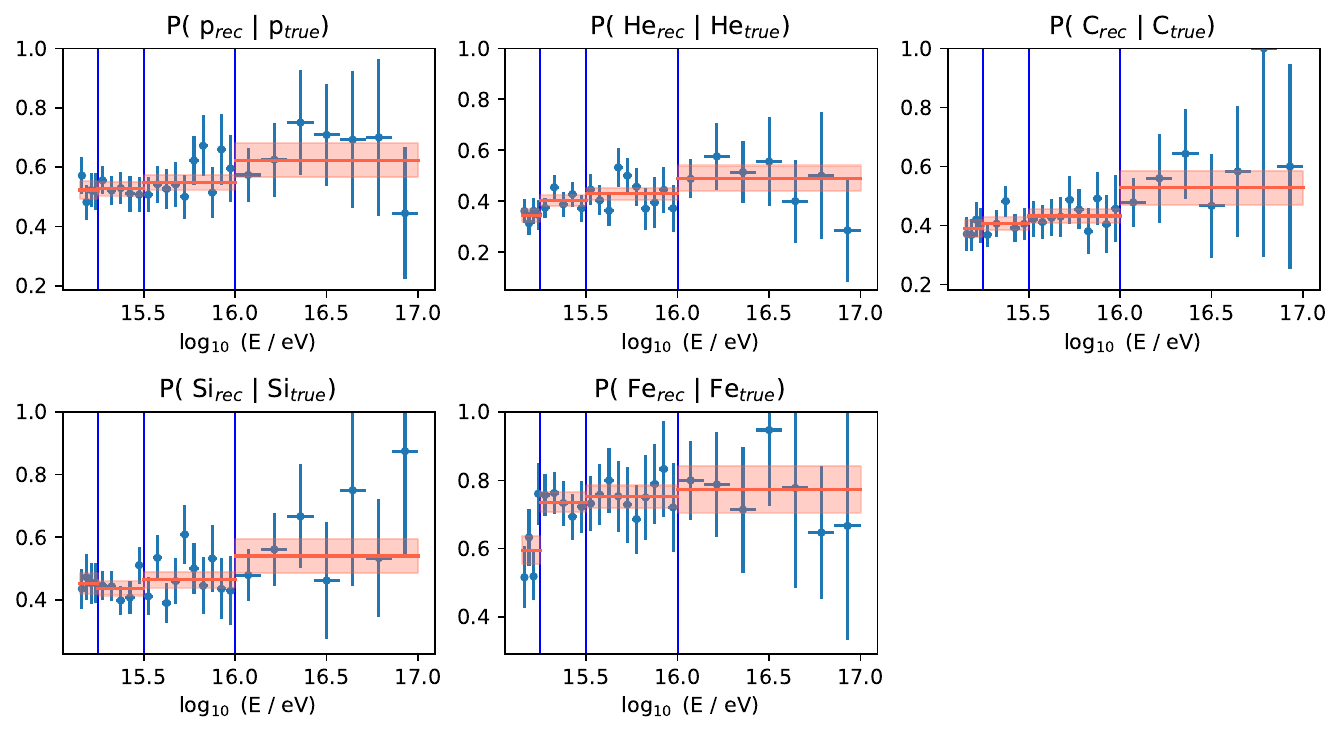}
    \caption{The dependence of the diagonal elements of the confusion matrix on the reconstructed energy for the CNN trained and tested using the \mbox{QGSJet-II.04} hadronic interaction model. 
    Blue vertical lines depict the ranges where the average confusion matrices are computed for the further unfolding procedure (see text). Red horizontal lines show the average value of a given matrix element in a given energy range together with its uncertainty.
    }
    \label{fig:diagonal confusion matrices energy bins}
\end{figure}

To figure out the evolution of the total accuracy of the mass components reconstruction with energy we look for the behavior of the diagonal elements of the confusion matrix, as they have the largest impact on the result. Their dependence on energy is shown in Fig.~\ref{fig:diagonal confusion matrices energy bins}. We split the full energy range under consideration into 4 subranges for each of which we separately compute the confusion matrix which is used for the further unfolding procedure (see Section~\ref{sec:unfold}). These ranges are: $15.15 < \log_{10} (E/\text{eV}) < 15.25$, $15.25 < \log_{10} (E/\text{eV}) < 15.5$, $15.5 < \log_{10} (E/\text{eV}) < 16$, $ \log_{10} (E/\text{eV}) > 16$. The chosen splitting has two goals. First, the matrix should be stable in the given energy range to avoid unnecessary uncertainties in the unfolding. Second, the overall accuracy of the mass component reconstruction, including the uncertainty from the limited Monte-Carlo statistics, should grow with energy.

\subsection{Cross-hadronic models reconstruction}
\label{sec:tests:hadr-mod}
\begin{figure}[t]
    \centering
    \begin{subfigure}{.32\textwidth}
        \includegraphics[width=\textwidth]{figs/cnn/final_redraw/cm_qgs.pdf}
       \caption{Train: QGS. Test: QGS}
    \end{subfigure}
    \hfill
    \begin{subfigure}{.32\textwidth}
        \includegraphics[width=\textwidth]{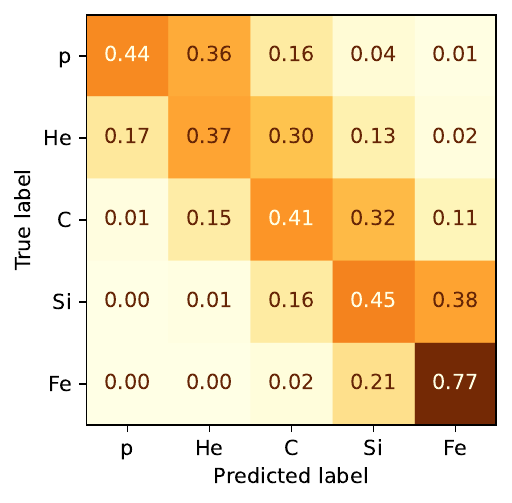}
       \caption{Train: QGS. Test: EPOS}
    \end{subfigure}
    \hfill
    \begin{subfigure}{.32\textwidth}
        \includegraphics[width=\textwidth]{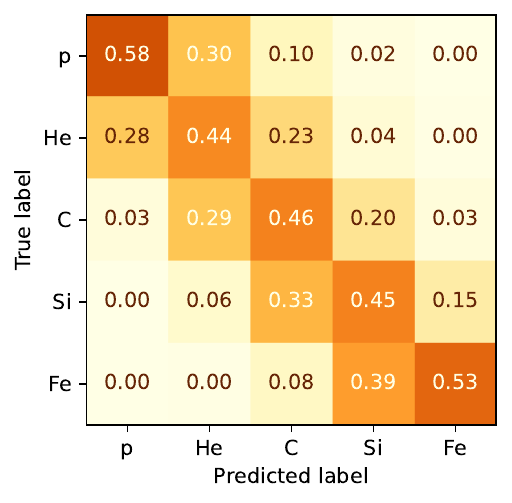}
       \caption{Train: QGS. Test: Sibyll}
    \end{subfigure}
    \vfill
    \begin{subfigure}{.32\textwidth}
        \includegraphics[width=\textwidth]{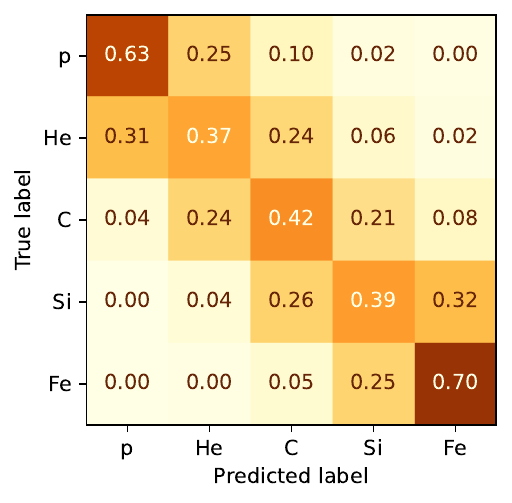}
       \caption{Train: EPOS. Test: QGS}
    \end{subfigure}
    \hfill
    \begin{subfigure}{.32\textwidth}
        \includegraphics[width=\textwidth]{figs/cnn/final_redraw/cm_epos.pdf}
       \caption{Train: EPOS. Test: EPOS}
    \end{subfigure}
    \hfill
    \begin{subfigure}{.32\textwidth}
        \includegraphics[width=\textwidth]{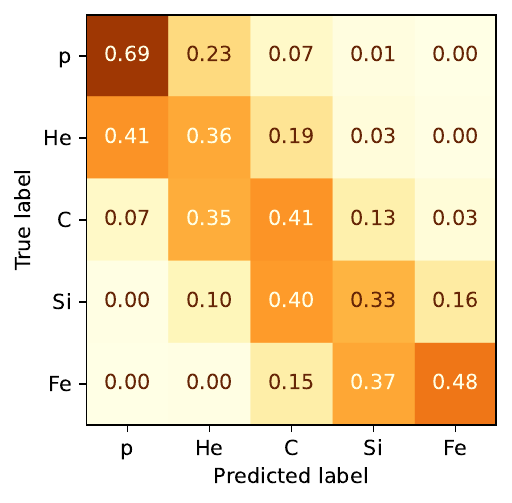}
       \caption{Train: EPOS. Test: Sibyll}
    \end{subfigure}
    \vfill
    \begin{subfigure}{.32\textwidth}
        \includegraphics[width=\textwidth]{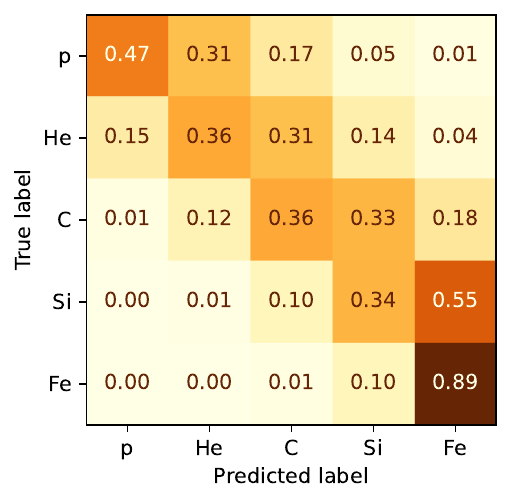}
       \caption{Train: Sibyll. Test: QGS}
    \end{subfigure}
    \hfill
    \begin{subfigure}{.32\textwidth}
        \includegraphics[width=\textwidth]{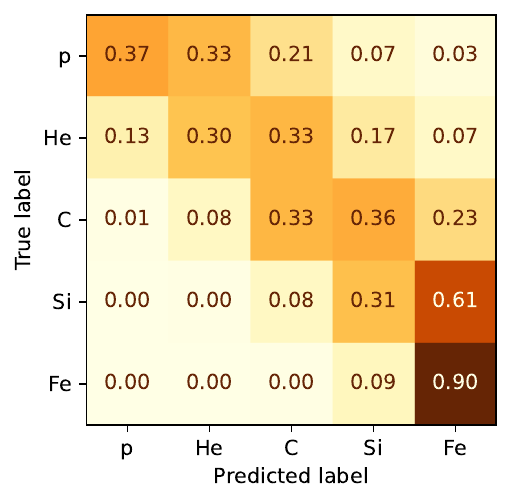}
       \caption{Train: Sibyll. Test: EPOS}
    \end{subfigure}
    \hfill
    \begin{subfigure}{.32\textwidth}
        \includegraphics[width=\textwidth]{figs/cnn/final_redraw/cm_sbl.pdf}
       \caption{Train: Sibyll. Test: Sibyll}
    \end{subfigure}
     
    \caption{Cross-hadronic model confusion matrices for the CNN models trained and tested separately with three different hadronic interaction models: QGSJet-II.04, EPOS-LHC and Sibyll\,2.3c which are marked in the figure as QGS, EPOS, and Sibyll, respectively. 
    }
    \label{fig:cnn cross hadronic}
\end{figure}

Here we consider the effect of the different hadronic interaction models. We trained and tested the CNN with the three modern hadronic interaction models: QGSJet-II.04, EPOS-LHC and Sibyll 2.3c. By comparing the predictions of the CNN trained on one interaction model and tested on another one, we can estimate how much our ML model is affected by their difference.
The resulting confusion matrices are shown in Fig.~\ref{fig:cnn cross hadronic}.
One can see that the predictions between QGSJet-II.04 and EPOS-LHC are generally similar and that the largest discrepancy is between Sibyll 2.3c and other models, where the notable shifts with respect to the main diagonal appear. At the same time, the general structure of the cross-hadronic model confusion matrices is smooth, without any sharp artifacts that give additional credibility to the ML model. 
Further in this study, we are not discussing the impact of the cross-hadronic model systematic on the resulting reconstruction of mass components spectra, leaving this issue for the full-scale mass composition analysis.

\subsection{Test with the unblind set of the real data}
\label{sec:tests:unblind}
\begin{figure}[t]
    \centering
    \includegraphics[width=\linewidth]{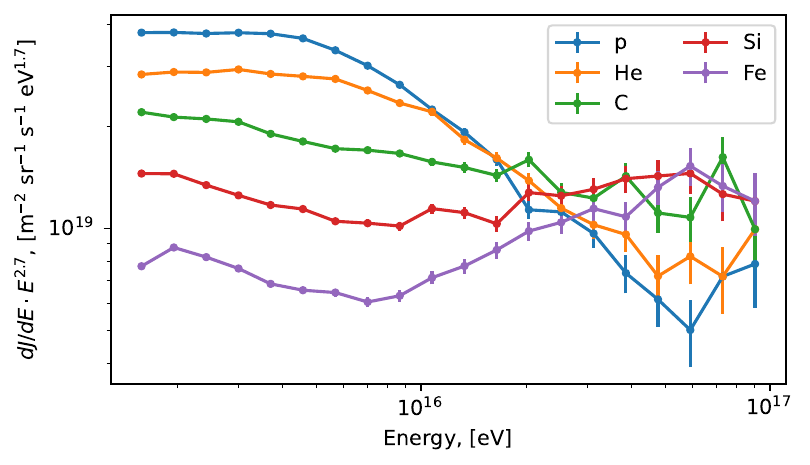}
    \caption{Mass composition (folded) spectra on unblind experimental data for the CNN model trained using QGSJet-II.04 hadronic interaction model.}
    \label{fig:unblind spectra cnn}
\end{figure}

To test our ML methods for possible hidden systematics that do not appear in Monte-Carlo, we use a so-called ``semi-blind'' data analysis in this study. 
As it was mentioned, we divide the experimental data into blind and unblind parts in a ratio of 80\%:20\%. Here we show the test of our ML reconstruction procedure with the unblind data set. In Fig.~\ref{fig:unblind spectra cnn} we show the spectra for the separate mass components of the unblind part of the KASCADE data according to the reconstruction of the CNN model trained with QGSJet-II.04 Monte-Carlo. 
One can see that the spectra of the separate mass groups are smooth at lower energies, where the statistical errors are small enough. The absence of apparent artifacts in these spectra increases the credibility of our mass composition analysis method.
We should note that for a correct comparison between these spectra and the original KASCADE spectra~\cite{Apel:2013uni} the unfolding effects should be taken into account and the hadronic model should be the same. In the next section, we build the CNN model trained with the QGSJet-II.02 model, perform the unfolding and make such a comparison.

\section{Unfolding and results}
\label{sec:unfold}
The accuracy of our methods of mass component reconstruction is not ideal. The same is true for the standard method of the primary energy reconstruction that we use in this study. At the same time, the uncertainties of the reconstructions are characterized by the confusion matrix of the given method for the given Monte-Carlo set. Using this knowledge we can improve the reconstruction for observables that are averaged over ensembles of events, for instance for spectra of separate mass components. This procedure of the reconstruction improvement is known as unfolding. In this section, we describe the method we use to unfold the reconstructed mass components spectra. We also compute the uncertainties that affect the final result and compare it to the original KASCADE reconstruction.

\subsection{Unfolding procedure}
\label{sec:perf:unfold}
We perform the unfolding for primary mass classification and for primary energy reconstruction separately, one after one. We start with the unfolding of a primary particle type. Let's denote the ratio of the number of events in $i$-th bin of some quantity to all events in a set as a probability $P_i$. For a single event, we have a standard formula of conditional probability:
\begin{equation}
\label{eq:unf_p1}
    P(A^{rec}_n | E^{rec}_m) = \sum_j P(A^{rec}_n | A^{tr}_j, E^{rec}_m) P(A^{tr}_j | E^{rec}_m)\, ; 
\end{equation}
where $A^{rec/tr}_i$ denotes $i$-th class of the primary particle (reconstructed or true) and
$E^{rec}_m$ denotes $m$-th bin of the reconstructed energy.
The same relation can be presented in another notation:
\begin{equation}
    Y_n^m = \sum_j R_{nj}^m X_j^m\, ;
\end{equation}
where $Y_n^m$ is a number of reconstructed particles of class~$n$ in the reconstructed energy bin~$m$; 
$R_{nj}^m$ is a so-called response matrix for the reconstructed energy bin~$m$ which is equal to the respective transposed confusion matrix
and $X_j^m$ is a number of true particles of class~$j$ in the reconstructed energy bin~$m$ that we want to derive.
This is a standard unfolding problem that can be solved in different ways.
Here we use a Bayesian iterative approach~\cite{DAgostini:1994fjx} with python package pyunfold~\cite{Bourbeau2018}. Stopping criteria $\Delta \chi^2$ are: $0.01$ for particle type unfolding and $0.1$ for energy unfolding. The uncertainties in response matrices were considered as Poissonian ones.

Then we continue with the energy unfolding. 
This can be described by the following relation:
\begin{equation}
\label{eq:unf_e1}
    P(E^{rec}_m) = \sum_i P(E^{rec}_m | E^{tr}_i) P(E^{tr}_i)\, ;
\end{equation}
where $E^{rec/tr}_m$ is $m$-th bin of reconstructed or true energy.
We also use the Bayesian iterative approach here with an initial suggestion of $P(E^{tr}_i)$ taken from the Monte-Carlo data.

Finally, we combine these two unfoldings.
The probability we need to know, $P(A^{tr}_i, E^{tr}_j)$, can be written as:
\begin{equation}
\label{eq:total unfolding}
    P(A^{tr}_i, E^{tr}_j) = P(A^{tr}_i | E^{tr}_j) P(E^{tr}_j) \to P(A^{tr}_i | E^{rec}_j) P(E^{tr}_j)\, ;
\end{equation}
here the arrow denotes that we substitute $P(A^{tr}_i | E^{tr}_j)$ with $ P(A^{tr}_i | E^{rec}_j)$ as this is the quantity we derive from the particle type unfolding. 
The impact of this substitution is considered as one of the systematics uncertainties of the method, it is estimated in the next Section using the Monte-Carlo data. Note, that the ``confusion matrices'' for the energy unfolding (of the energy reconstruction with Eq.~\ref{eq:e_rec}) are computed separately in each energy bin, while that for the particle type unfolding are averaged over energy intervals defined in the Section~\ref{sec:perf:unfold}.

\subsection{Estimation of the uncertainties}
\label{sec:tests:unfold-uncert}
Here we estimate the total uncertainty of our reconstruction of mass components spectra, in particular, we take into account statistical uncertainty associated with a limited number of experimental events and a number of systematic uncertainties:
\begin{itemize}
    \item uncertainty due to missing detectors, as described in Sec.~\ref{sec:tests:miss-det};
    \item uncertainty of the response matrix related to a limited amount of Monte-Carlo data;
    \item uncertainty related to a bias due to the unfolding regularization;
    \item uncertainty of the energy unfolding due to a different energy resolution for different mass components
    \item uncertainty of the full unfolding procedure due to the substitution in Eq.~\ref{eq:total unfolding}.
    \item uncertainty related to a spectrum index in the Monte-Carlo
    \item uncertainty related to non-full experiment efficiency at low energies
\end{itemize}
The total systematic uncertainty is calculated as a sum in quadrature of all separate systematic uncertainties.
Since the latest KASCADE results are based on the QGSJet-II.02 hadronic interaction model, we are using the same model to make a valid comparison.
We do not include the uncertainties related to the hadronic models in the present analysis, leaving it for a further full-scale KASCADE composition study. All the systematic uncertainties described in this Section are calculated in a way that is appropriate for both Monte-Carlo sets and experimental data sets, where the true values of parameters are unknown.
The statistical uncertainty of the unfolded mass components spectra is calculated by propagation of the standard statistical uncertainty with the pyunfold package.

The uncertainty due to missing detectors is estimated as follows. 
First, we evaluate the response matrices for the default Monte-Carlo without missing detectors and the Monte-Carlo imitating missing detectors (see Section~\ref{sec:tests:miss-det}) and find their difference.
Then, we use this difference as the uncertainty of the response matrix and propagate it into the uncertainty of the resulting mass component spectra.
The resulting uncertainty is taken with the weight of $1/2$, the fraction of events with missing detectors in the experimental data.
In terms of a flux in a given energy bin, the relative value of this uncertainty is $1-20\%$ on average for all mass components, depending on the primary particle energy.
The impact of the missing detectors on the energy unfolding is evaluated using the data from the unblind set. We construct two sets: one with all detectors working in all events, and another one with a number of detectors non-working in all events. The difference in the energy unfolded all-particle spectra of these two sets is interpreted as the uncertainty due to the missed detector. This uncertainty is $\sim 4\%$ on average for all energies.

There is a statistical uncertainty in the response matrix coming from a limited number of Monte-Carlo events. We propagate this into a systematic uncertainty of the unfolded mass components spectra using pyunfold package. The relative value of this uncertainty for particle type unfolding in the resulting spectra is $6-20\%$ on average for all mass components, depending on the energy bin. The same relative uncertainty for energy unfolding is $2-15\%$.

The unfolding method itself is not ideal, it contains an iterative procedure with predefined stopping parameters, that leads to the appearance of a bias in the unfolded results. This bias appears in both the particle type unfolding and the energy unfolding.
Theoretically, we need a very large amount of Monte-Carlo events to estimate this bias. In practice, we do the estimations with the following steps.
We name the raw ML predictions of component fractions as (vector) $\nu$. Then we unfold these predictions using our procedure and consider the resulting values, $\mu$, as true ones. Then we fold the results back, just by multiplying them with the confusion matrices, and name the result $\tilde{\nu}$. The discrepancy between $\nu$ and $\Tilde{\nu}$ describes the non-ideality of our unfolding procedure. To translate this discrepancy into the uncertainty of the reconstructed spectra we produce 100 toy Monte-Carlo spectra, randomized using Poisson PDF with mean $\tilde{\nu}$, and unfold them again using our procedure. We name the obtained fractions: $\tilde{\mu}$. Then the resulting uncertainty is estimated as difference $\langle \tilde{\mu} \rangle - \mu$ in each energy bin.
The described procedure has the advantage that the uncertainty estimated for the data set has no extra contributions due to possible correlations of the unfolding bias with other uncertainties in Monte-Carlo.
For the particle type unfolding the relative value of this uncertainty in the resulting spectra is $1-65\%$ ($1-11\%$ if we do not account for bins where we have outliers, see Fig~\ref{fig:qgs02 compare}) on average for all mass components, depending on the energy bin. For the energy unfolding the respective uncertainty is $0.1-13\%$, depending on the energy bin.

We also estimated the uncertainty in the energy unfolding associated with different energy resolutions for mass groups.
For example, the energy response matrix for \emph{Fe} is better than for \emph{p}.
Thus, if the given mixture in a Monte-Carlo set used for energy response matrix computation is different from the one we have in a data set, additional uncertainty will appear.
To estimate uncertainty we perform two extra energy unfolding separately, using response matrices for \emph{p} only and for \emph{Fe} only. Then we consider the mix obtained after the particle type unfolding and assign the proton energy unfolding to \emph{p} and \emph{He} components but iron energy unfolding for all heavier components, for simplicity. 
We compare the resulting spectra with those where energy unfolding was performed with a standard response matrix.
This uncertainty has an $8-10\%$ value depending on energy.

\begin{figure}[t]
    \centering
    \includegraphics[width=\linewidth]{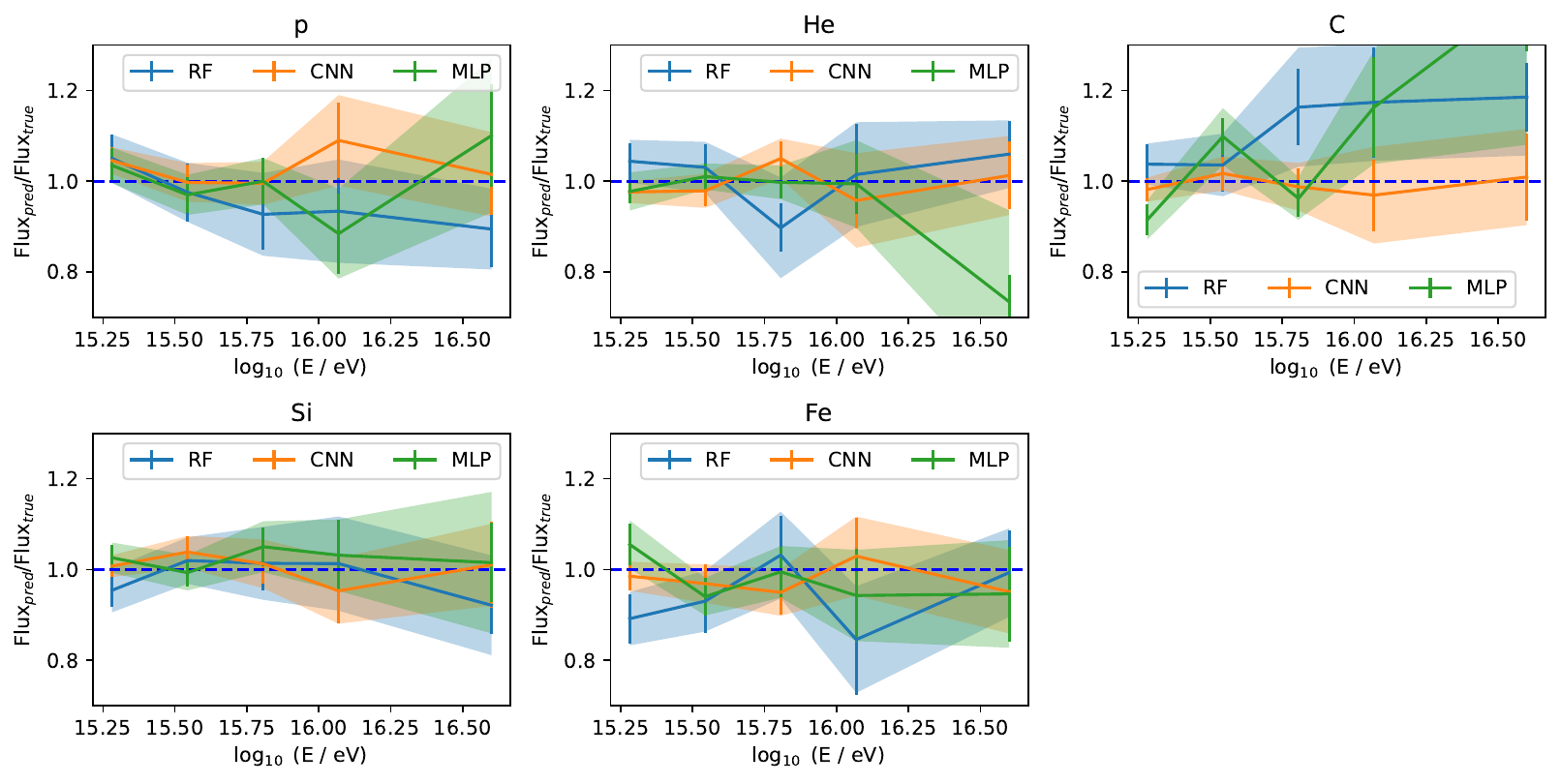} 
    \caption{Predicted spectra for each mass component after particle type unfolding performed with the different ML models. Monte-Carlo set with QGSJet-II.04 hadronic interaction model is used.
    Whiskers represent statistical errors. Bands represent systematic uncertainty that includes the estimated bias of the unfolding procedure and the uncertainty of the response matrix.
    }
    \label{fig:particle unfolding mc}
\end{figure}

\begin{table}[t]
    \centering
\begin{tabular}{|l||*{2}{c|}}\hline
\backslashbox{Uncertainty}{Unfolding type}
&\makebox[6em]{Particle type}&\makebox[6em]{Energy}
\\\hline\hline
Missing detectors  & $1-20\%$ & $4\%$ \\\hline
Limited MC & $6-20\%$ & $2-15\%$ \\\hline
Unfolding regularization & $1-11 (65)\%$ & $0.1-13\%$ \\\hline
Substitution $P(A^{tr}_i | E^{tr}_j) \rightarrow P(A^{tr}_i | E^{rec}_j)$ & $0-13\%$ & - \\\hline
Energy resolution in mixes & - & $8-10\%$ \\\hline
Energy spectrum index in MC & - & $0.5-7\%$ \\\hline
Detection \& cuts efficiency & - & $0-8\%$ \\\hline\hline
Total ($13 - 30 (79)\%$) & $6-24 (73)\%$ & $11-20\%$ \\\hline
\end{tabular}
    \caption{Summary of relative values of systematic uncertainties for CNN reconstruction of the mass components spectra. Monte-Carlo set with QGSJet-II.02 hadronic interaction model is used. All the uncertainty values are averaged over all mass components, the range of the values shown is an uncertainty variation within the energy range studied. The values in brackets include the uncertainty of the regularization bias of the particle type unfolding, computed including outlier bins. For missing detectors, the uncertainty of the energy unfolding is given as an average overall energy range studied. Total uncertainty corresponds to a sum of all uncertainties in quadrature.}
    \label{tab:uncert}
\end{table}

The uncertainty from Eq.~\ref{eq:total unfolding} was estimated with Monte-Carlo sets. First, we perform both particle type unfolding and energy unfolding, fit the resulting points using $\chi^2$ procedure, and consider the resulting spectra as ``true'' values. Then we fold these spectra using the energy response matrix derived from the Monte-Carlo. This yields us spectra as it would be reconstructed by our ML methods from the given ``true'' spectra. Therefore, the ratio of these spectra in a given energy bin yields a sought for uncertainty:
\begin{equation}
    1 + \delta = \frac{P(A^{tr}_i | E^{tr}_j)}{P(A^{tr}_i | E^{rec}_j)} 
\end{equation}
The value of this uncertainty is growing from almost zero at low energies up to 13\% at high energies, on average for all mass components.
This behavior is explained by the fact that the mass group fractions change dramatically around the primary particle energy $10^{16}$~eV, where the \emph{Fe} fraction rises and the \emph{p} fraction falls.
So in the region of the largest derivative of the fractions, the uncertainties reach maximum values.  

We also take into account uncertainty due to possible differences in a spectral index $\gamma$ between the Monte-Carlo set and the real data set. This uncertainty appears because the response matrix used for energy unfolding depends on $\gamma$ assumed in the MC set.
To estimate this effect we perform the energy unfolding with $\gamma = -3.0$ response matrix and calculate the difference from our basic result with $\gamma = -2.7$. This yields an extra uncertainty of $0.5-7\%$ depending on the energy of the primary particle.

There is an extra uncertainty related to our procedure of spectrum reconstruction. Ideally, the spectra for each mass component should be computed taking into account the incomplete efficiency of detection and quality cuts for this component at a given energy.
While in our procedure the spectra for all components are computed assuming the full efficiency. This leads to an appearance of extra uncertainty in low-energy bins. We estimate the uncertainty related to this assumption by calculating the 
difference between the spectrum reconstructed with our method and the one reconstructed assuming the efficiency averaged over all components in a given energy bin. There is an $8\%$ uncertainty for the first bin, a $1\%$ for the second bin, and no extra uncertainty for the higher energy bins. The summary of all calculated uncertainties is given in Table~\ref{tab:uncert}.

The test of the particle type unfolding with the Monte-Carlo sets for three different ML methods is shown in Fig.~\ref{fig:particle unfolding mc}. 
We divided the Monte-Carlo test sets into two equal parts. The first part was used for the response matrix computation, while the second part~--- for the spectra unfolding. Note, that the results presented in this Figure are based on the Monte-Carlo set with QGSJet-II.04 hadronic interaction model, while the general uncertainties calculations presented in this paragraph were based on the Monte-Carlo set with QGSJet-II.02 model.
From this figure, one can derive two conclusions. First, the predictions of all three ML methods have a general agreement with the spectra of the true components, within their uncertainties. Second, there is also a good agreement between the unfolded predictions of different ML methods. Both of these findings provide additional credibility to our unfolding method.

\subsection{Comparison with the standard KASCADE reconstruction}
\label{sec:KASCADE_comparison}
\begin{figure}[t]
    \centering
    \includegraphics[width=\linewidth]{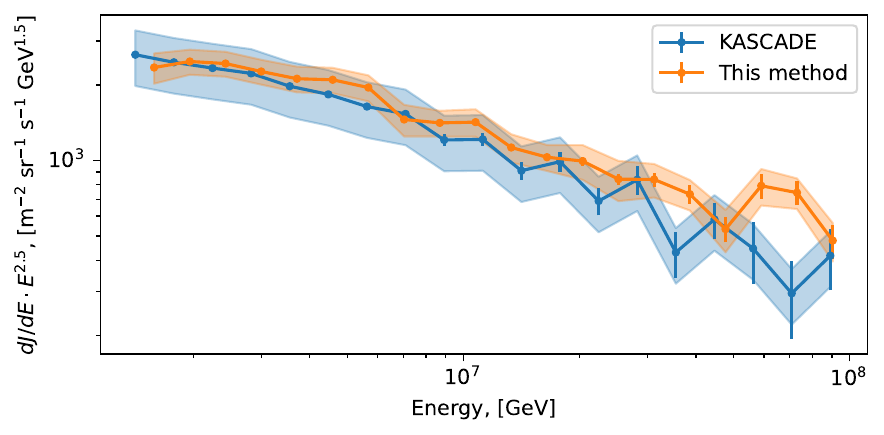}
    \caption{Comparison of all-particle spectra with energy unfolding for our reconstruction and the original KASCADE reconstruction. Energy estimation is based on the QGSJet-II.02 hadronic interaction model in both spectra. 
    Whiskers show the statistical uncertainties. Bands denote systematic uncertainties (see text).
    }
    \label{fig:energy unfolding}
\end{figure}

\begin{figure}[t]
    \centering
    \includegraphics[width=\linewidth]{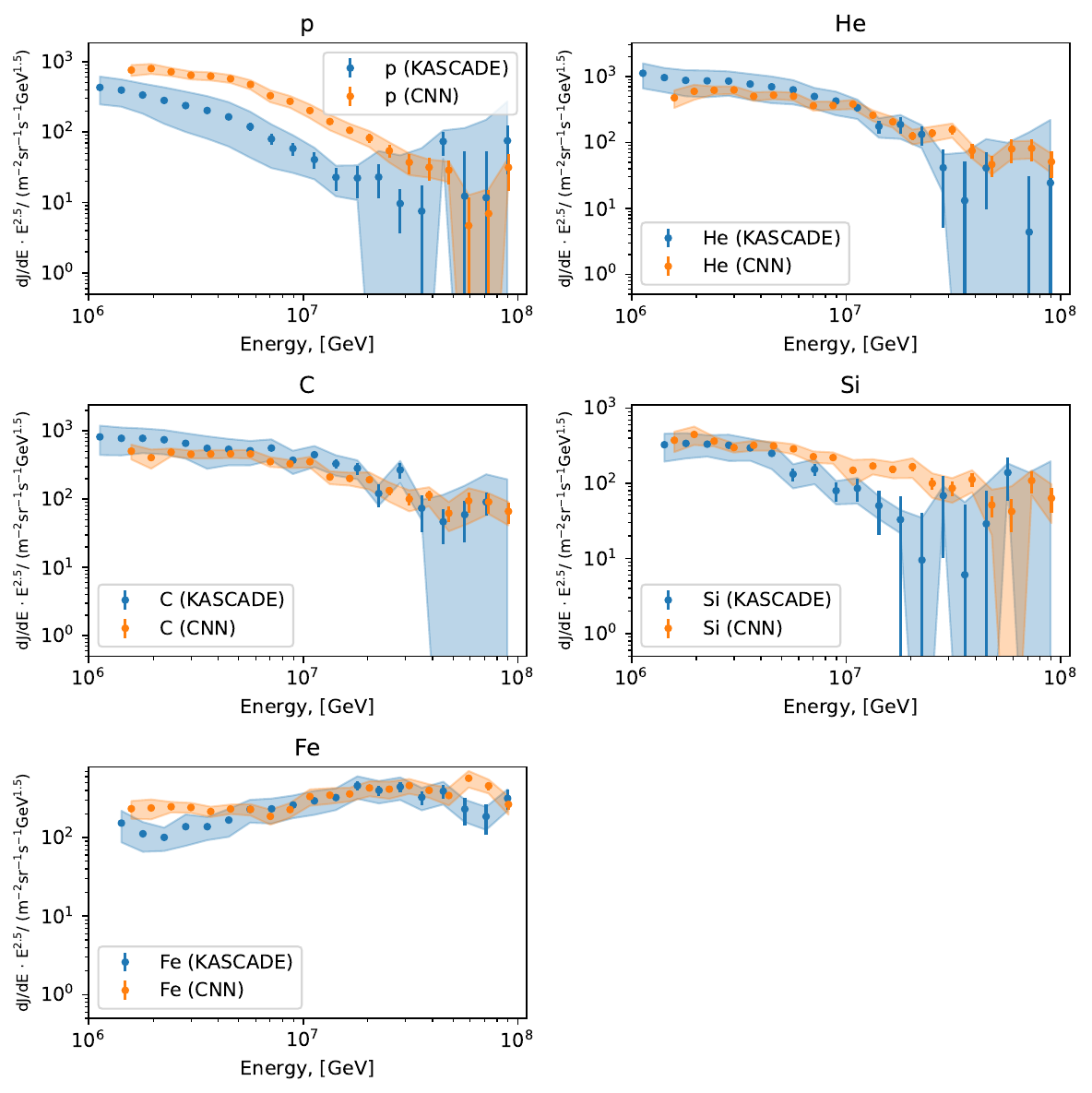}  
    \caption{Comparison of the mass components spectra for the experimental data reconstructed using the original KASCADE methods~\cite{Apel:2013uni} and using our CNN method. QGSJet-II.02 hadronic interaction model is used in both cases.}
    \label{fig:qgs02 compare}
\end{figure}

In this section, we compare our results for the unblind part of experimental data with the original KASCADE results of mass components spectra reconstruction~\cite{kascade_cuts, Apel:2013uni, Finger_2011}.
First, we present the result of our reconstruction of the all-particle spectrum. It includes the energy unfolding and all uncertainties related to the energy reconstruction, as described in Section~\ref{sec:tests:unfold-uncert}.
Namely, the uncertainties of the finite response matrix, missing detectors impact, impact of mixtures, efficacy, different $\gamma$ of the MC total flux ($-2.7$ and $-3.0$), and the unfolding method are included.
The comparison of these results with the original KASCADE all-particle spectrum is shown in Fig.~\ref{fig:energy unfolding}.
One can see that both spectra are in agreement within the uncertainties in almost all studied energy range. However, our spectrum has smaller uncertainties, around $11-20\%$, than $\sim 25\%$ uncertainty of the original KASCADE spectrum. Note, that our statistical uncertainties are smaller since we use the larger data set: our unblind set contains $\sim 1.6 \cdot 10^6$ events after quality cuts (without $\lg (E/eV) > 15.15$ cut), while the set used in the latest KASCADE analysis~\cite{Finger_2011} contains only $\sim 6.4 \cdot 10^5$ events. Note also, that our all-particle spectrum reconstruction does not use any ML methods. The minor discrepancy in the high energy part of the spectra can happen because of the non-smoothness of the energy unfolding in these bins, that in turn originates from the small amount of Monte-Carlo events in these bins and resulting fluctuations of the confusion matrices from bin to bin: the uncertainty due to unfolding bias could be somewhat underestimated for these bins. This situation is characteristic for both our method and the original KASCADE reconstruction.

Finally, we present the comparison of separate mass component spectra reconstructed using our CNN method with that of the original KASCADE method. Again, the unblind part of the experimental data set is used for our reconstruction.
We perform an unfolding and take into account all the uncertainties described in Section~\ref{sec:tests:unfold-uncert}. Our method is trained and tested with the QGSJet-II.02 hadronic interaction model, to make the comparison valid.
The results are shown in Fig.~\ref{fig:qgs02 compare}. One can see that the uncertainties of our method are smaller than those of the original KASCADE method. Nevertheless, the spectra for all components except protons are generally in agreement with the uncertainties.

\section{Conclusions}
\label{sec:conclusion}
In this study, we have developed several new machine learning methods for the analysis of the public data of the KASCADE experiment. Our goal was the reconstruction of the energy spectra of separate mass components of cosmic ray flux in the energy range from 1 to 100 PeV. We have tested the performance of 4 different ML methods built for mass components classification: random forest, multi-layer perceptron, convolutional neural network, and EfficientNet. All the ML methods were trained and tested with the Monte-Carlo data of the KASCADE experiment. The best accuracy and the most stable performance were demonstrated by the CNN model. We made several cross-checks and stability tests for this network, including tests with a small part of the real data, the unblind set. All the tests were passed successfully proving the stability of the CNN. We also performed the unfolding of the reconstructed mass components spectra and estimated various uncertainties related to this procedure and to Monte-Carlo in general. The resulting all-particle and component-wise spectra built with the unblind data set were compared to that from the original composition analysis of the KASCADE experiment. The overall accuracy of our reconstruction methods was shown to significantly exceed that of the original KASCADE method for both all-particle and component-wise spectra.

\acknowledgments
We are grateful to Dmitry Kostunin for the inspiration of this study and for the assistance at its various stages.
We would like to thank Grigory Rubtsov, Ivan Kharuk, Vladimir Lenok and Victoria Tokareva for the fruitful discussions and comments.
The work was supported by the Russian Science Foundation grant 22-22-00883.

\begin{appendices}
\section{Details of the neural networks architectures}
\label{app:arch}
\begin{figure}[t]
    \centering
    \includegraphics[width=0.7\textwidth, angle=0]{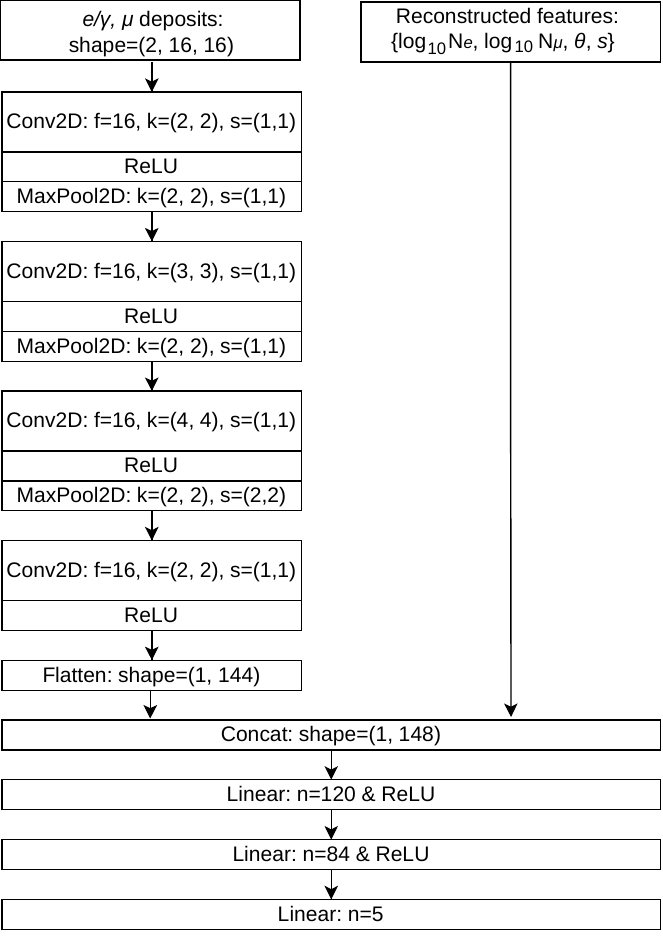}
    \caption{Architecture of the convolutional neural network. Data flow is represented by arrays. 
    Layer names are given as in PyTorch.
    Abbreviations: ``f''~-- the number of filters, ``k''~-- kernel size, ``s''~-- strides, ``n''~-- the number of neurons.}
    \label{cnn_architecture}
\end{figure}
\begin{figure}[t]
    \centering
    \includegraphics[width=0.6\textwidth, angle=0]{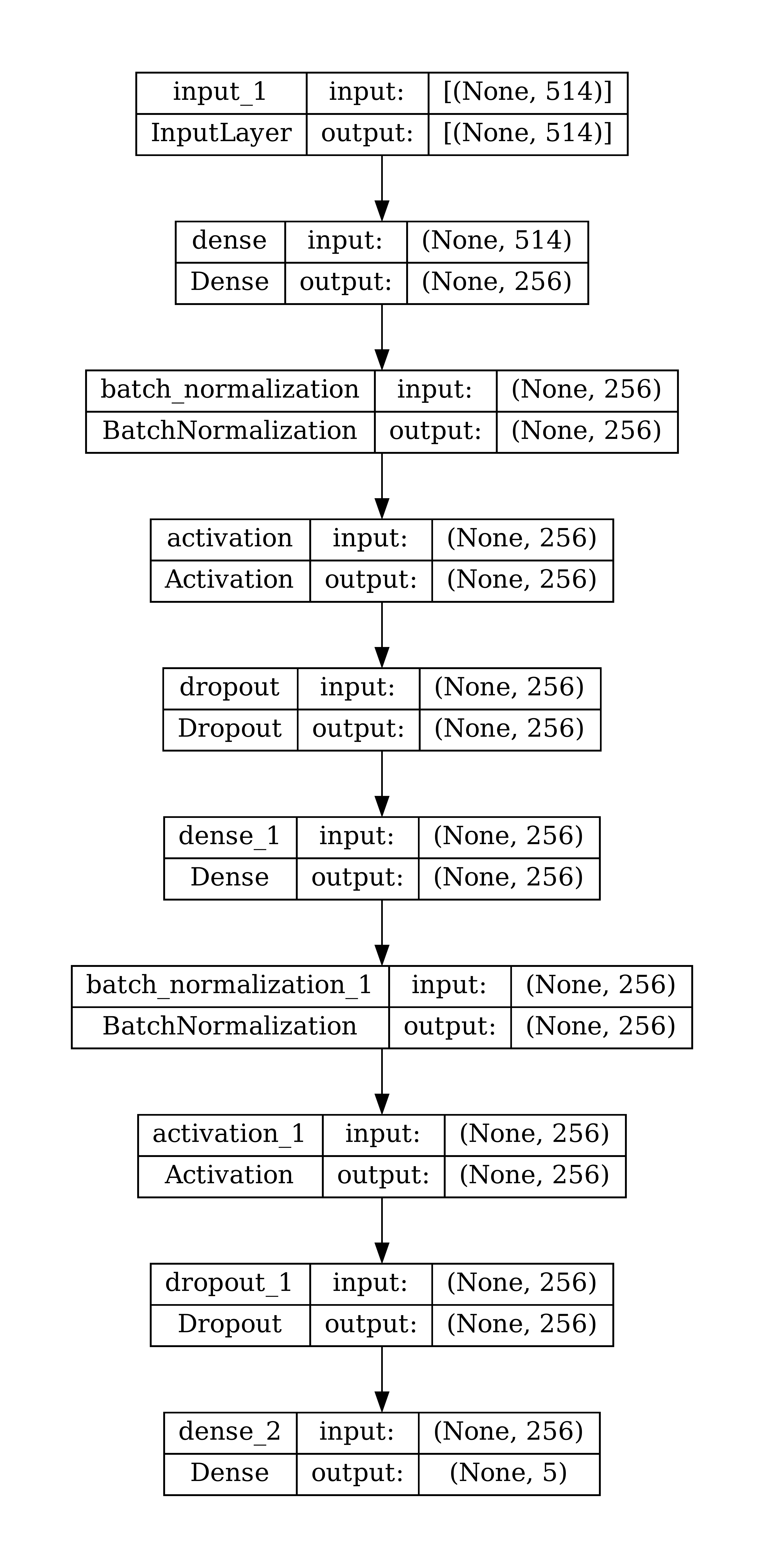}
    \vspace{-1cm}
    \caption{Architecture of the feed-forward neural network (MLP). The flow of data is shown by arrays. 
    Layer names are given in TensorFlow format. ``None'' in tensor shapes means variable batch size shape.}
    \label{mlp_architecture}
\end{figure}

This section presents details of the implemented neural network models. 
In particular, the architecture of the CNN model is shown in Fig.~\ref{cnn_architecture} and that of the MLP model is shown in Fig.~\ref{mlp_architecture}. For both CNN and MLP models, we first pre-process the data before feeding it to the input of the classifier. The values of deposits are rescaled to zero-mean and unit-variance normal distribution $\mathcal{N}(0, 1)$, and reconstructed features are min-max scaled into the range $[0, 1]$.
Minor data augmentation, such as event image rotation, is also used to make the training process more stable. 

The architecture of the CNN model consists of two main parts. 
The first part is convolutional filters. We apply a set of these filters to deposits from $e/\gamma$ and $\mu$ detectors. This part is designed to identify the patterns corresponding to a particular primary mass component.
The second part is dense layers. It supplies the results of the first part with reconstructed high-level EAS event features and combines them for the classification of the mass components. 
Therefore, the input of the CNN model consists of deposits from $e/\gamma$ and $\mu$ detectors represented as $16\times16$ image with 2 channels and high-level features: $\log_{10} N_e$, $\log_{10} N_{\mu}$, $\theta$, $s$.
The CNN classifier is implemented in PyTorch~\cite{PyTorch}. The model has $\sim$\,$36\,000$ trainable parameters. We use Adam optimizer~\cite{kingma2014adam} for training this model and Weights\,\&\,Biases package~\cite{wandb} for tracking of the training process.

\begin{figure}[t]
    \centering
    \begin{subfigure}{.32\textwidth}
        \includegraphics[width=\textwidth]{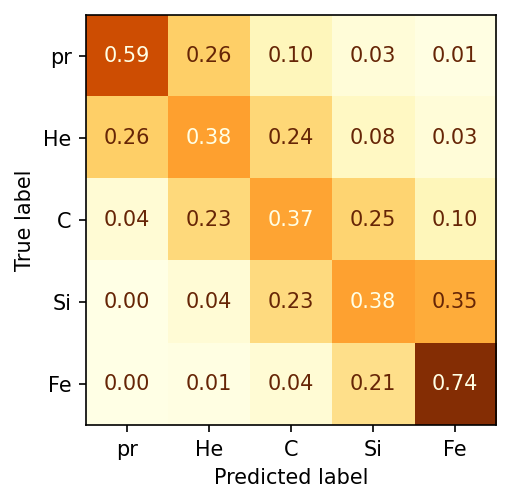}
       \caption{Train: QGS. Test: QGS}
    \end{subfigure}
    \hfill
    \begin{subfigure}{.32\textwidth}
        \includegraphics[width=\textwidth]{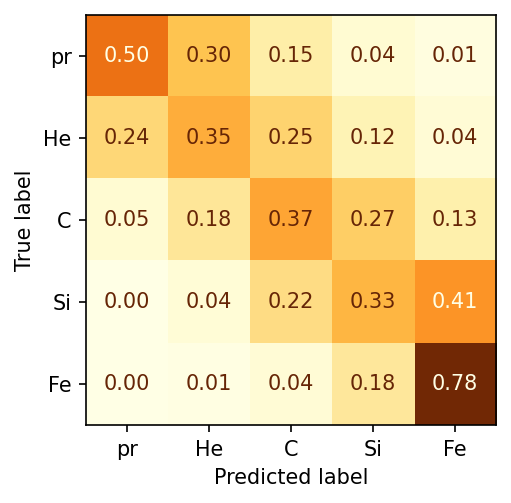}
       \caption{Train: QGS. Test: EPOS}
    \end{subfigure}
    \hfill
    \begin{subfigure}{.32\textwidth}
        \includegraphics[width=\textwidth]{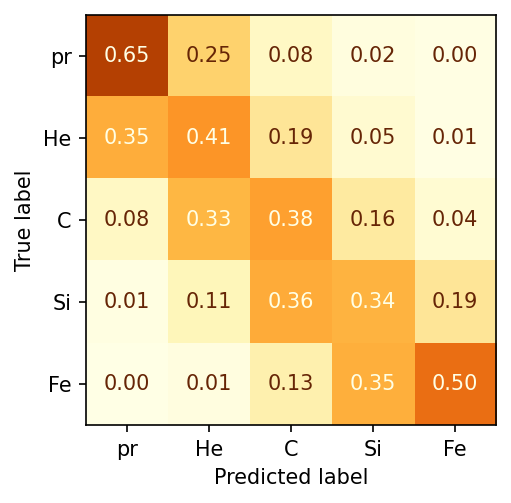}
       \caption{Train: QGS. Test: Sibyll}
    \end{subfigure}
    \vfill
    \begin{subfigure}{.32\textwidth}
        \includegraphics[width=\textwidth]{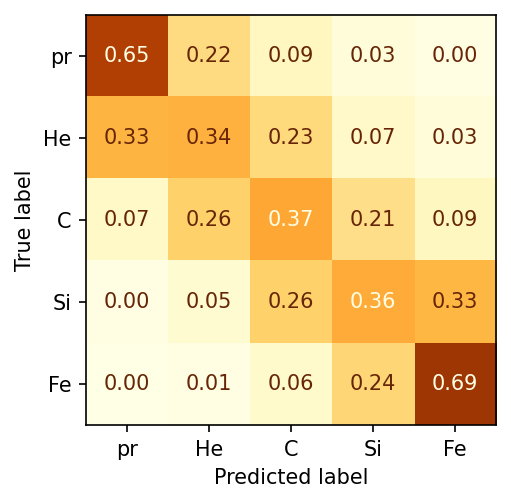}
       \caption{Train: EPOS. Test: QGS}
    \end{subfigure}
    \hfill
    \begin{subfigure}{.32\textwidth}
        \includegraphics[width=\textwidth]{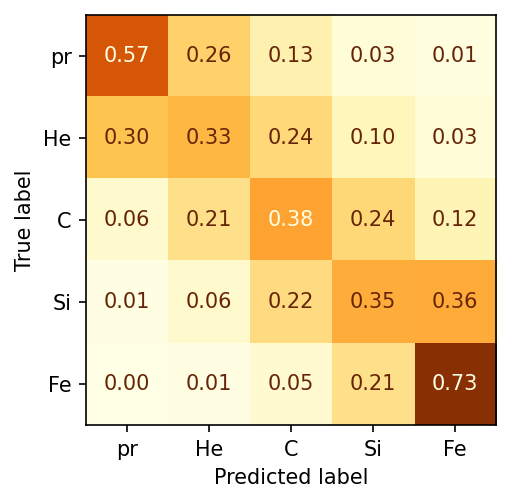}
       \caption{Train: EPOS. Test: EPOS}
    \end{subfigure}
    \hfill
    \begin{subfigure}{.32\textwidth}
        \includegraphics[width=\textwidth]{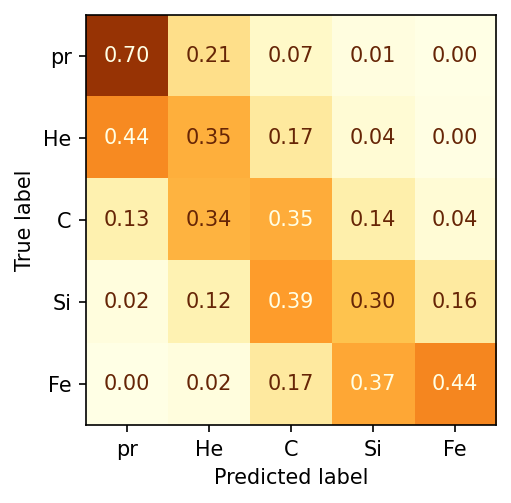}
       \caption{Train: EPOS. Test: Sibyll}
    \end{subfigure}
    \vfill
    \begin{subfigure}{.32\textwidth}
        \includegraphics[width=\textwidth]{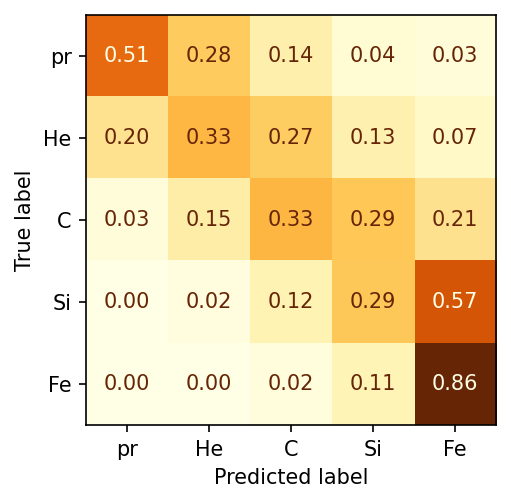}
       \caption{Train: Sibyll. Test: QGS}
    \end{subfigure}
    \hfill
    \begin{subfigure}{.32\textwidth}
        \includegraphics[width=\textwidth]{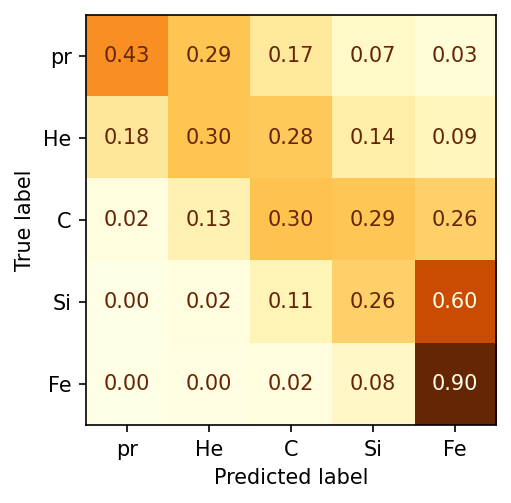}
       \caption{Train: Sibyll. Test: EPOS}
    \end{subfigure}
    \hfill
    \begin{subfigure}{.32\textwidth}
        \includegraphics[width=\textwidth]{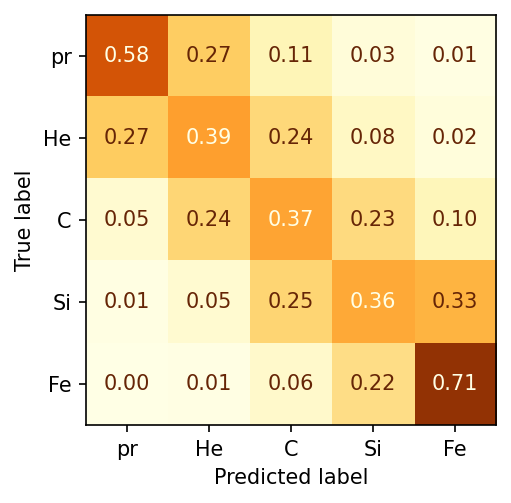}
       \caption{Train: Sibyll. Test: Sibyll}
    \end{subfigure}
    \caption{Cross-hadronic model confusion matrices for the RF classifier models trained and tested separately with three different hadronic interaction models: QGSJet-II.04, EPOS-LHC and Sibyll\,2.3c which are marked in the figure as QGS, EPOS, and Sibyll, respectively.
}
    \label{fig:rfcls_hadr_models_comparison}
\end{figure}

The MLP model's architecture is extremely simple: it consists of only two dense (feedforward) layers, each followed by batch normalization, ELU activation, and dropout with a rate of $0.15$. The MLP model uses a slightly smaller number of inputs compared to the CNN model: it similarly deposits from $e/\gamma$ and $\mu$ detectors (although this time they're converted to 1-dimensional flat array format) and only two high-level features: $\theta$ and $\phi$. The model has $\sim$\,$200\,000$ trainable parameters and is implemented using TensorFlow~\cite{abadi2016tensorflow} and Keras~\cite{chollet2015keras}. 

\section{Details and tests for Random Forest model}
\label{app:rf-tests}
\begin{figure}[t]
    \centering
    \begin{subfigure}{.32\textwidth}
        \includegraphics[width=\textwidth]{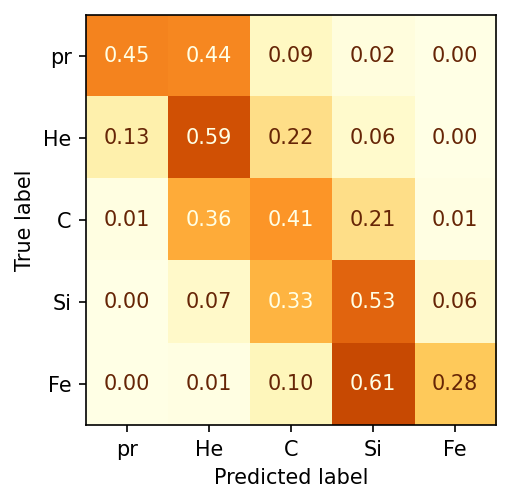}
       \caption{Train: QGS. Test: QGS}
    \end{subfigure}
    \hfill
    \begin{subfigure}{.32\textwidth}
        \includegraphics[width=\textwidth]{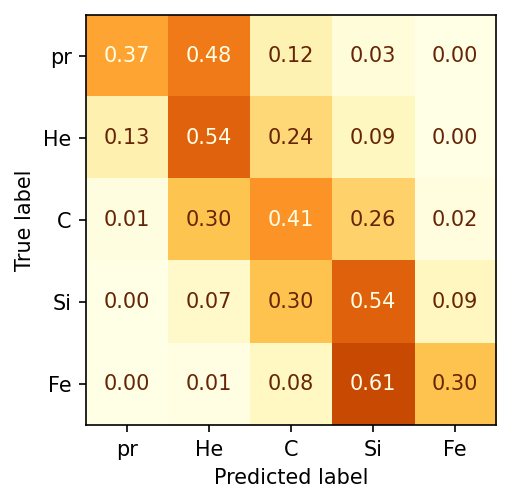}
       \caption{Train: QGS. Test: EPOS}
    \end{subfigure}
    \hfill
    \begin{subfigure}{.32\textwidth}
        \includegraphics[width=\textwidth]{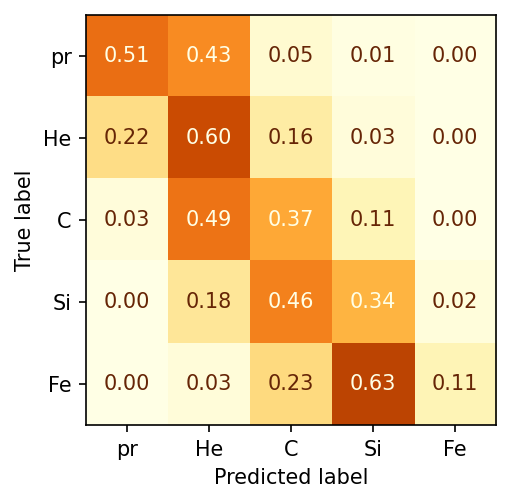}
       \caption{Train: QGS. Test: Sibyll}
    \end{subfigure}
    \vfill
    \begin{subfigure}{.32\textwidth}
        \includegraphics[width=\textwidth]{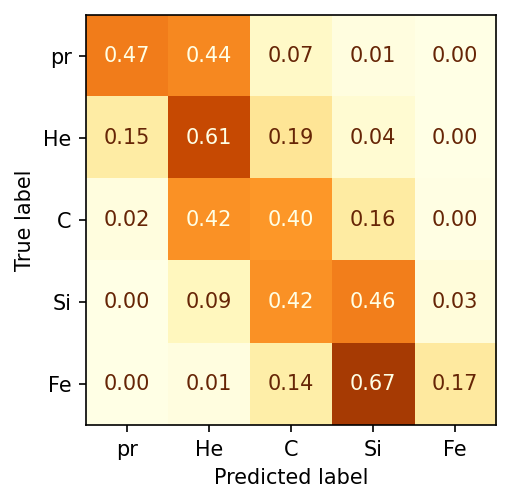}
       \caption{Train: EPOS. Test: QGS}
    \end{subfigure}
    \hfill
    \begin{subfigure}{.32\textwidth}
        \includegraphics[width=\textwidth]{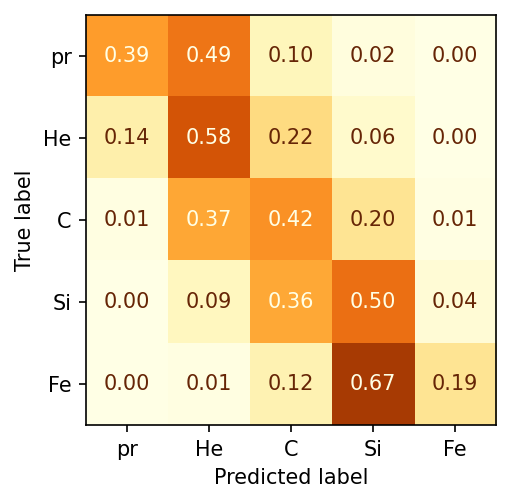}
       \caption{Train: EPOS. Test: EPOS}
    \end{subfigure}
    \hfill
    \begin{subfigure}{.32\textwidth}
        \includegraphics[width=\textwidth]{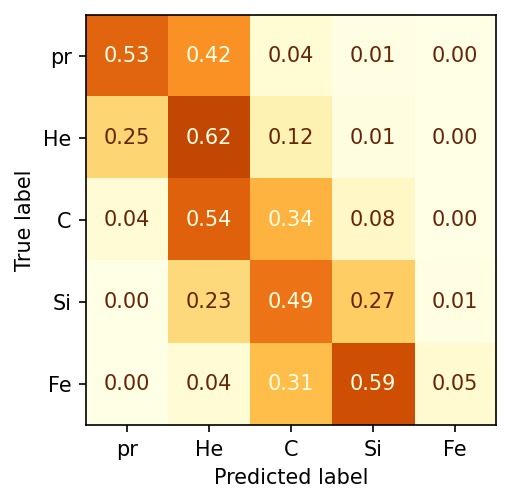}
       \caption{Train: EPOS. Test: Sibyll}
    \end{subfigure}
    \vfill
    \begin{subfigure}{.32\textwidth}
        \includegraphics[width=\textwidth]{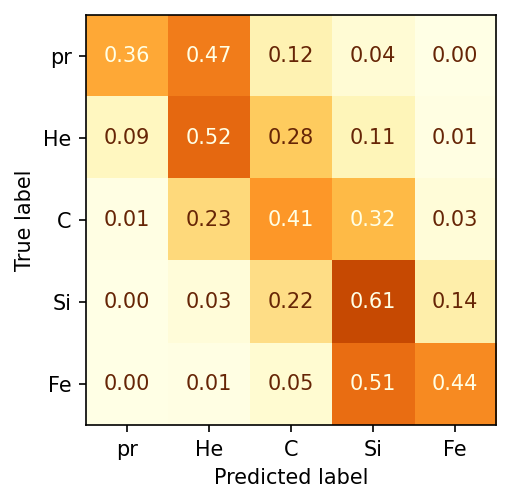}
       \caption{Train: Sibyll. Test: QGS}
    \end{subfigure}
    \hfill
    \begin{subfigure}{.32\textwidth}
        \includegraphics[width=\textwidth]{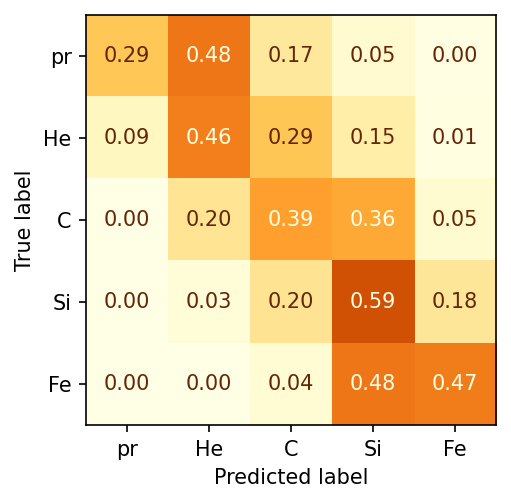}
       \caption{Train: Sibyll. Test: EPOS}
    \end{subfigure}
    \hfill
    \begin{subfigure}{.32\textwidth}
        \includegraphics[width=\textwidth]{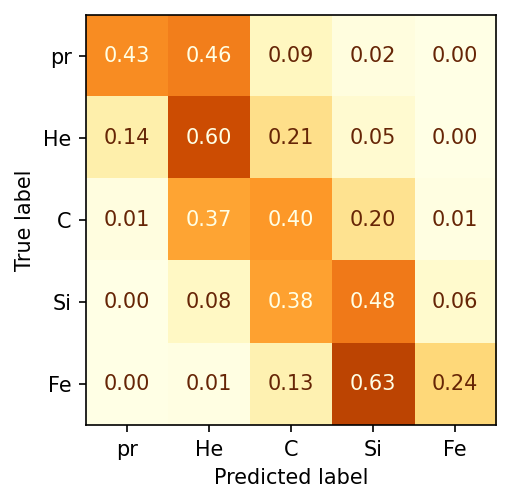}
       \caption{Train: Sibyll. Test: Sibyll}
    \end{subfigure}
    \caption{Cross-hadronic model confusion matrices for the RF regressor models trained and tested separately with three different hadronic interaction models: QGSJet-II.04, EPOS-LHC and Sibyll\,2.3c which are marked in the figure as QGS, EPOS, Sibyll, respectively.}
    \label{fig:rfreg_hadr_models_comparison}
\end{figure}

Here we present the details for the RF model optimization and show some tests performed for this model. 
The following parameters were optimized for the RF classifier:
\begin{itemize}
\item The number of estimators (trees): [1, 10, 100, 500, 1000],
\item The maximum depth of the tree: [5, 10, 50, 100, 'not limited'],
\item Split quality criteria: ['gini', 'entropy'],
\item Maximum number of features to consider when looking for the best split: ['num features', 'sqrt(num features)'],
\item Class weights: ['equal', 'balanced'].
\end{itemize}
In Fig.~\ref{fig:rfcls_hadr_models_comparison} we show the reconstruction of cross-hadronic models for the RF classifier. One can see that the performance of this classifier for cross-hadronic model reconstruction is close to that of CNN, with a little less accuracy for intermediate mass components.

Apart from the RF classifier model the RF regressor model was designed.
The following parameters were optimized for this model:
\begin{itemize}
\item The number of estimators (trees): [1, 10, 100, 500, 1000],
\item The maximum depth of the tree: [5, 10, 50, 100, 'not limited'],
\item Split quality criteria: ['squared error', 'absolute error'],
\item Maximum number of features to consider when looking for the best split: ['num features', 'sqrt(num features)'].
\end{itemize}

In Fig.~\ref{fig:rfreg_hadr_models_comparison} we show the reconstruction of cross-hadronic models for the RF regressor. As one can see, this ML model shows a significant off-diagonality of the confusion matrices, especially in the ``light'' and ``heavy'' parts of the matrix. Importantly, this problem persists even if the regressor is trained and tested with one and the same hadronic interaction model. This makes the given ML model unsuitable for the mass composition reconstruction in the given setup.
\end{appendices}

\suppressfloats

\bibliography{ref}

\end{document}